\documentclass[sigconf]{acmart}
\AtBeginDocument{%
  }

\copyrightyear{2026}
\acmYear{2026}
\setcopyright{cc}
\setcctype{by}
\acmConference[ICSE '26]{2026 IEEE/ACM 48th International Conference on Software Engineering}{April 12--18, 2026}{Rio de Janeiro, Brazil}
\acmBooktitle{2026 IEEE/ACM 48th International Conference on Software Engineering (ICSE '26), April 12--18, 2026, Rio de Janeiro, Brazil}
\acmPrice{}
\acmDOI{10.1145/3744916.3773258}
\acmISBN{979-8-4007-2025-3/26/04}

\usepackage{enumitem}
\usepackage{multirow}
\usepackage{algorithmic}
\usepackage{subfig}
\usepackage{textcomp}
\usepackage{url}
\usepackage{listings}
\usepackage{pifont}
\usepackage{colortbl}
\usepackage{xspace}

\newlist{questions}{enumerate}{2}
\setlist[questions,1]{label=\textbf{RQ\arabic*:},ref=RQ\arabic*}
\setlist[questions,2]{label=(\alph*),ref=\thequestionsi(\alph*)}

\newcommand{\rev}[1]{\textcolor{black}{#1}}

\newcommand{\RedL}{\makebox[0pt][l]{\color{red!6}\rule[-3pt]{\linewidth}{10pt}}}
\newcommand{\GreenL}{\makebox[0pt][l]{\color{green!6}\rule[-3pt]{\linewidth}{10pt}}}
\definecolor{green}{RGB}{0,127,0}

\lstdefinelanguage{diff}{
  basicstyle=\scriptsize\ttfamily \color{black},
  columns=fullflexible,
  breaklines=true,
  breakatwhitespace=false,
  showspaces=false,               
  showstringspaces=false,  
  frame=single, 
  showtabs=false,
  numbersep=5pt,
  showstringspaces=false,        
  stepnumber=1,                   
  tabsize=5,                     
  title=\lstname,  
  numbers=left,                 
  numbersep=5pt,   
  backgroundcolor=\color{white},
  morecomment=[f][\lstbg{red!5}]-,
  morecomment=[f][\lstbg{green!5}]+,
  morecomment=[f][\textit]{@@},
}
\usepackage{fontawesome}
\newcommand{\toolName}{PyQu\xspace}
\newcommand{\totalCommits}{3.7 million\xspace}
\newcommand{\totalLoC}{2.7 trillion\xspace}
\newcommand{\datasetSize}{15,597\xspace}
\newcommand{\changes}{61\xspace}
\newcommand{\categories}{13\xspace}

\begin{document}

\title{From Code Changes to Quality Gains: An Empirical Study in Python ML Systems with PyQu}

\author{Mohamed Almukhtar}
\email{almukhtr@umich.edu}
\affiliation{%
  \institution{University of Michigan-Flint}
  \country{USA}
}
\author{Anwar Ghammam}
\email{aghammam@umich.edu}
\affiliation{%
  \institution{University of Michigan-Dearborn}
  \country{USA}
}
\author{Marouane Kessentini}
\email{kessentm@gvsu.edu}
\affiliation{%
  \institution{Grand Valley State University}
  \country{USA}
}
\author{Hua Ming}
\email{huaming@umich.edu}
\affiliation{%
  \institution{University of Michigan-Flint}
  \country{USA}
}


\begin{abstract}
In an era shaped by Generative Artificial Intelligence for code generation and the rising adoption of Python-based Machine Learning systems (MLS), software quality has emerged as a major concern. As these systems grow in complexity and importance, a key obstacle lies in understanding exactly how specific code changes affect overall quality—a shortfall aggravated by the lack of quality assessment tools and a clear mapping between ML systems code changes and their quality effects. Although prior work has explored code changes in MLS, it mostly stops at what the changes are, leaving a gap in our knowledge of the relationship between code changes and the MLS quality.

To address this gap, we conducted a large-scale empirical study of 3,340 open-source Python ML projects, encompassing more than 3.7 million commits and 2.7 trillion lines of code. We introduce PyQu, a novel tool that leverages low-level software metrics to identify quality-enhancing commits with an average accuracy, precision, and recall of 0.84 and 0.85 of average F1 score. Using PyQu and a thematic analysis, we identified 61 code changes, each demonstrating a direct impact on enhancing software quality, and we classified them into 13 categories based on contextual characteristics. 41\% of the changes are newly discovered by our study and have not been identified by state-of-the-art Python changes detection tools. Our work offers a vital foundation for researchers, practitioners, educators, and tool developers, advancing the quest for automated quality assessment and best practices in Python-based ML software.
\end{abstract}






\maketitle

\vspace{-0.7em}
\section{Introduction}
\label{sec:Introduction}

In recent years, advancements in Generative AI for code generation, coupled with the widespread adoption of Machine Learning (ML) techniques across a broad range of fields, have dramatically transformed modern software development workflows \cite{pandey2024transforming,habehh2021machine, dixon2020machine}. While systems leveraging these technologies achieve impressive levels of accuracy across diverse domains \cite{dheepak2021comprehensive}, they also grow increasingly complex and operate at large scales \cite{cote2024quality}. As prior research points out, ensuring high-quality Machine Learning Systems (MLS) is essential for their long-term maintainability, reliability, and overall effectiveness in real-world applications\cite{jabborov2023taxonomy, sommerville2011software, fenton2014software}.

 Despite the importance of maintaining high-quality MLS in general, and Python-based in specific—due to its dominance in both industry and open-source MLS \cite{braiek2018open, dilhara2021understanding, raschka2019python}—identifying which code changes genuinely improve the quality of Python-based MLS remains a challenge. Moreover, assessing these improvements on a large scale necessitates automated tools, which are often unavailable. While previous studies \cite{Dilhara_Ketkar_Sannidhi_Dig_2022, tang2021empirical} have examined the types of changes commonly applied in MLS, to the best of our knowledge, no research has systematically analyzed their impact on overall \rev{ML} systems quality.

 Understanding code changes that enhance ML systems helps developers make informed decisions to improve and maintain their software effectively. To illustrate the impact of code changes on ML system quality, we present an example in \autoref{lst:MotEx}. The author moves model configuration from the class constructor into a separate file (\texttt{Constant.py}). Initially, the constructor accepts a configuration parameter (line 1) and saves it in a class field (line 3).

 Now, it is replaced with CategoricalParameter(ACTIVATIONS) (line 4), to retrieve the configuration externally.
 This change enhances maintainability by centralizing parameters for easier modification and improves modularity by separating concerns, enabling reuse without duplication.
This example illustrates a frequent type of code modification in Python-based MLS which improves the quality of the code, yet has not been investigated in prior research on Python code changes \cite{tang2021empirical}, nor identified by state-of-the-art changes detection tools \cite{pyref, PyRefMiner, Dilhara_Ketkar_Sannidhi_Dig_2022}. This brings broader questions to light: \rev{How frequently Python-specific quality-enhancing changes occur—particularly those overlooked by existing tools, and which software Quality Attributes (QAs) do they impact?}

  \begin{center} 
\begin{lstlisting}[language=diff,caption={Commit \textit{78bd9c13} in danielhers/tupa: Externalize model configuration.},label={lst:MotEx},numbers=left,frame=lines, escapechar=!, ,columns=fixed]
!\RedL!def __init__(self, args, params, init, dropout, activation):
!\GreenL!def __init__(self, args, params):
!\RedL!   self.activation = activation
!\GreenL!   self.activation = CategoricalParameter(ACTIVATIONS) 
\end{lstlisting} 
\end{center}


To tackle these challenges, we conducted a large-scale, empirical investigation on code changes that enhance quality in open-source Python MLS. We specifically target traditional, code-centric ML systems—where pipelines are built, trained, and maintained directly in source code using libraries such as TensorFlow and PyTorch.   Our dataset includes 3,340 repositories, over \totalCommits\ commits, and more than \totalLoC\ lines of code, making it, to our knowledge, the most extensive study of its kind focused on quality-related code changes for Python-based MLS and their impact on the software quality. As part of this effort, we developed \toolName\, a novel tool designed to detect quality-enhancing commits. 
\rev{Unlike traditional tools that assign subjective, absolute labels such as “understandable code” or “maintainable code”~\cite{Scalabrino2017, Scalabrino_Linares‐Vásquez_Oliveto_Poshyvanyk_2018, sjoberg2012questioning},} \toolName\ measures low-level software metrics before and after a change and \rev{employ ML classification models} to provide a more objective assessment of whether the quality has enhanced or not. \toolName\ achieved an average accuracy, precision, and recall of 0.84, with an F1-score of 0.85, demonstrating its effectiveness and generalizability in identifying meaningful quality improvements.

Leveraging \toolName\ to detect quality-enhancing commits, along with a thematic analysis, we identified \changes\ distinct types of code changes that directly improve different MLS \rev{QAs}, categorized based on their contextual behavior into \categories\ categories. Out of these changes, 25 are newly discovered, neither explored in previous research on Python code changes~\cite{tang2021empirical} nor detected by state-of-the-art changes detection tools~\cite{pyref, PyRefMiner, Dilhara_Ketkar_Sannidhi_Dig_2022}. 
In this context, we address the following research questions:


\begin{questions}

\item \textbf{How effective is \toolName\ in detecting quality-enhancing commits in MLS?}
This RQ aims to evaluate the tool’s effectiveness in detecting quality-enhancing commits in Python-based MLS, using a ground-truth dataset of manually annotated commits, and standard evaluation metrics. \toolName\ achieved an average accuracy, precision, and recall of 0.84 and 0.85 F1-score.

\item \textbf{How well does \toolName\ generalize to unseen code changes?} 
This RQ aims to evaluate \toolName\ generalizability by testing it on unseen code changes. Our tool demonstrated high generalizability indicated by the tight train-test accuracy $\leq$ 0.05, and a high ROC-AUC score $\geq$ 0.84 across different \rev{QAs}.

\item \textbf{What are the quality-enhancing code changes in MLS?} 
This RQ aims to build a taxonomy of code changes that improve MLS \rev{QAs}. We were able to develop a taxonomy of \changes\ change types classified into \categories\ main categories and mapped them to the specific \rev{QAs} they improve.\newline

\end{questions}

The main contributions of this paper are:
\begin{enumerate}
    \item We introduce \toolName, a novel tool that uses low-level quality metrics to assess Python-based MLS code quality. To encourage further research, we make \toolName\ publicly available \cite{MLCodeQuality}.


    

    \item To the best of our knowledge, this is the first large-scale study to systematically identify quality-enhancing code changes in Python-based MLS. We identified \changes\ types of code changes, mapping each to the \rev{QAs} they improve. Notably, \rev{25} were previously undocumented, highlighting key opportunities to enhance the state-of-the-art Python change detection tools. 

    \item We introduce a benchmark dataset to assess the quality of Python-based MLS \cite{MLCodeQuality}. It comprises 1,728 manually verified commits, each annotated with the \rev{QA} it enhanced. The dataset also includes the corresponding code changes for each commit, facilitating deeper analysis and further research.
    \end{enumerate}
\vspace{-0.7em}
\vspace{-0.7em}
\section{Methodology}
\label{sec:Methodology}

\begin{figure*}[h!]    
\centering
\includegraphics[width=0.9\textwidth]{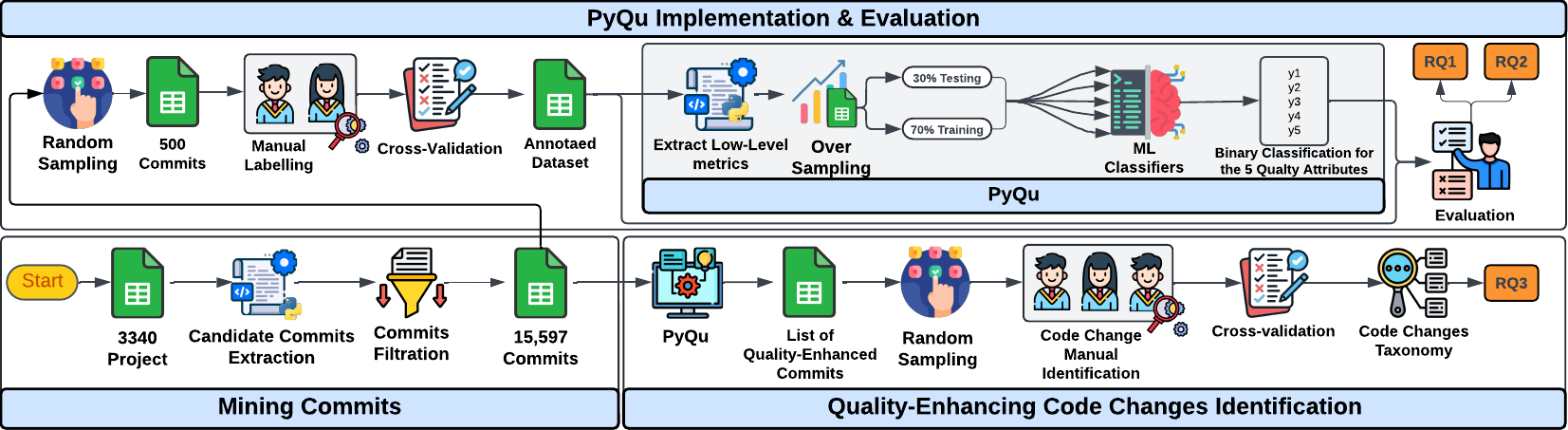}
\caption[Research Methodology Overview.]{Research Methodology Overview.}
\label{fig:methodologyOverview.}
\vspace{-0.4cm}
\end{figure*}

In this section, we introduce our systematic approach to analyzing code changes that impact the quality of Python-based MLS. 
\autoref{fig:methodologyOverview.} provides an overview of the process which is composed of three primary phases: 1) Mining Commits, 2) PyQu Implementation \& Evaluation, and 3) Quality-Enhancing Changes Identification. 

\vspace{-1em}
\subsection{Subject Systems}

\rev{We based our study on the repository list curated by Dilhara et al.~\cite{dilhara2021understanding}, comprising 3,340 top-rated open-source ML projects with over 3.7 million commits and 2.7 trillion lines of code. This list offers a diverse and representative sample of Python-based ML repositories across various sizes, contributor bases, and popularity levels. While we did not curate the repositories list, we mined and filtered its commit history—using commit message analysis, ML-related filtering, and file count constraints—to construct a refined dataset focused on software quality evolution. Although the original projects list was compiled in 2021, many of them remain active (e.g., TensorFlow~\cite{Tensorflow}), and we analyzed more recent versions to ensure the extracted dataset reflects current development practices. The commit mining and dataset construction process is detailed in the following sections.}


\vspace{-1em}
\subsection{Mining Commits}
\label{subsec:ExtractingQualityEnhancingCommits}



In this subsection, we describe our approach to constructing a dataset of commits intended to enhance software quality.
Our data extraction methodology  consists of two main steps: 


\vspace{-0.5em}
\subsubsection{Candidate Commits Extraction}


The first step for building our dataset involves extracting \rev{commits where developers explicitly indicated an intention to improve quality. This helps lower false positives from incidental changes and ensures we capture a comprehensive set of relevant commits.} 
To achieve this, we rely on commit messages and descriptions, a technique widely adopted in prior research to accurately reflect code change intent \cite{sliwerski2005changes,tang2021empirical}.

We adapted general terms from prior studies \cite{tang2021empirical}, and expanded them for broader inclusion. We selected different keywords such as `refactor', `enhance code', `improve code', `code quality', and `quality metric` to ensure extensive coverage. Furthermore, we applied regular expressions to the root keywords (e.g., ‘refactored’ and ‘refactoring’ from ‘refactor’) to capture all relevant variations and strengthen our list. 

\vspace{-0.2cm}
\subsubsection{Commits Filtration}

The commits filtration involves a two-step process: \textbf{\textit{1. ML-related Filtration:}} 
Our initial manual verification revealed that some commits did not involve ML-related files. \rev{Therefore, we refined our selection criteria by including only commits where the modified files explicitly import common ML libraries such as Keras, PyTorch, TensorFlow, Scikit-learn, NumPy, or Pandas~\cite{ShiriHarzevili2022, Nguyen2020, Sarkar2018}.} \textbf{\textit{2. File Count Filtration:}}
\rev{Since commits often span multiple files, we applied a conservative filter, including only those that modify five or fewer Python files following prior research~\cite{vitale2023using}. This reduces the risk of unrelated changes and increases the focus and relevance of observed modifications.}



The commits extraction pipeline yielded 15,597 commits that met all the specified conditions, serving as the foundation for our study.

\vspace{-1em}

\subsection{Studied Quality Attributes}
\label{subsec:formulas}

To assess the quality of the studied commits, we utilize 5 widely used QAs, including  Understandability, Reliability, Maintainability, Usability, and Modularity. In this subsection, we explore the specifics of these \rev{QAs} in our study and the rationale behind their selection. Our choice of the five \rev{QAs} is guided by both empirical evidence and prior research \cite{jabborov2023taxonomy,chen2019assessing,Scalabrino2019Automatically}. Jabborov et al. \cite{jabborov2023taxonomy} highlight the important role of reliability, maintainability, modularity, and usability as critical quality assessment \rev{attributes} for ML-based software systems. 
Other \rev{attributes} discussed by Jabborov et al. conceptually overlap with our selected \rev{QAs}, reducing the need for their separate inclusion. Moreover, understandability is a fundamental aspect of system design, as emphasized by Chen et al. and Scalabrino et al. \cite{chen2019assessing, Scalabrino2019Automatically}, influencing code readability, developer productivity, and long-term maintainability. 

By focusing on these five \rev{QAs}, we ensure that our study remains both rigorous and practically applicable, leveraging well-established low-level metrics to systematically evaluate the impact of code changes on Python-based MLS. The following subsections provide a detailed discussion of each selected quality and the associated low-level metrics used to quantify them. \rev{A comprehensive summary of these low-level metrics, including what they measure and how they are computed, is provided in~\autoref{tab:pyqu-metrics}.}
\vspace{-0.4em}
\subsubsection{\textbf{Understandability (UN)}} 
\label{subsub:understandability}
\rev{refers to how easily developers comprehend a system’s structure, logic, and behavior~\cite{Scalabrino2017, fenton2014software}. Higher understandability reduces cognitive load and supports knowledge sharing and onboarding~\cite{Scalabrino2019Automatically, scalabrino2016improving}. We assess understandability using a combination of low-level metrics that are widely adopted in prior research and collectively offer a strong foundation for Understandability's automated assessment\cite{mccabe1976complexity, ebert2016cyclomatic, DeYoung1982Analyzer-generated, scalabrino2016improving, Scalabrino_Linares‐Vásquez_Oliveto_Poshyvanyk_2018}. These include complexity metrics such as Cyclomatic Complexity (CC) and Halstead Volume (HV), where a \emph{decrease} generally corresponds to simpler control flow and improved understandability~\cite{mccabe1976complexity, Posnett2011A, DeYoung1982Analyzer-generated, ebert2016cyclomatic, Scalabrino2019Automatically}. We also consider documentation-related metrics such as Comment Readability (CR) and the Comment-to-Code Ratio (CCR), where \emph{higher values} indicate better commenting practices that aid in code comprehension~\cite{DeYoung1982Analyzer-generated, scalabrino2016improving, aljedaani2024boring, Scalabrino_Linares‐Vásquez_Oliveto_Poshyvanyk_2018, Scalabrino2019Automatically}. Additionally, stylistic conformance indicators contribute to evaluating code understandability. These include Style Conformance Score (SCS), where higher values indicate stronger adherence to style guidelines \cite{hariprasad2017style, PEP8}, Advanced Feature Penalty (AFP), where lower values reflect reduced use of obscure or advanced Python constructs \cite{Scalabrino2019Automatically}, and API/Framework Conformance (APIFC), where higher scores denote more consistent use of APIs and frameworks aligned with official best practices~\cite{EffectiveTensorflow2, PyTorchTutorials, khan2023apifc}—all of which have been shown to support understandability assessment.} \rev{We model (UN) as a function of changes ($\Delta$) in the low-level metrics, as defined in \autoref{eq:under_equation}. Each $\Delta$ term represents the difference in a metric before and after a commit (e.g., $\Delta SCS = SCS_{\text{post}} - SCS_{\text{pre}}$), capturing the impact of the change. As described in Section~\ref{subsub:classifycommits}, we train ML classifiers (MLC) on these deltas to predict whether a commit improves understandability.}

\vspace{-1.3em}
\begin{align}
\text{UN} = \text{MLC}\big(\Delta CC, \Delta HV, \Delta CCR, \Delta CR, \Delta SCS, \Delta APIFC, \Delta AFP \big)
\label{eq:under_equation}
\end{align}

\vspace{-1.3em}


\rev{
\begin{table*}[]
\begin{tabular}{|c|p{5cm}|p{11cm}|}
\hline
\multicolumn{1}{|c|}{\textbf{\rev{Met}}} & \multicolumn{1}{c|}{\textbf{\rev{What it Measures}}} & \multicolumn{1}{c|}{\textbf{\rev{Computation Method}}}                                                                              \\ \hline
\rev{CC}                                    & \rev{Control-flow complexity}& \rev{Calculated using the Radon library ~\cite{radon_2023} by analyzing control-flow graphs to count decision points.}                                                                       \\ \hline
\rev{HV }                                   & \rev{Cognitive complexity}& \rev{Calculated using Radon ~\cite{radon_2023} based on the number and types of operators and operands.}                                                                                       \\ \hline
\rev{LoC }                                  & \rev{Code size (Lines of Code)  }                                 & \rev{Counted using Radon library \cite{radon_2023}. }                                                                                                              \\ \hline
\rev{CCR   }                                & \rev{Comment density} & \rev{Ratio of comment lines to code lines. Both are extracted using Radon~\cite{radon_2023}. }                                                                                                 \\ \hline
\rev{CR}                                    & \rev{Comments Readability} & \rev{Computed using Textstat library \cite{textstat_2022} for docstrings and comments readability, based on Flesch–Kincaid readability score \cite{Scalabrino2019Automatically}.                   }                                                                                 \\ \hline
\rev{SCS  }                                 &\rev{Styling Compliance Score}                 & \rev{Computed using Flake8~\cite{Flake8} and Flake8-Tensor~\cite{flake8-tensors_2024} libraries based on the PEP-8 rules \cite{PEP8}}                                                                                                       \\ \hline
\rev{AFP  }                                 & \rev{Number of advanced/less readable Python features (e.g., lambdas) }   & \rev{Violations are detected using custom checkers built on Python's AST parser \cite{ast}. }                                                                                      \\ \hline
\rev{APIFC}                                 & \rev{Correct use of ML framework APIs score}               & \rev{Checked using Custom Pylint rules conformance with TensorFlow/PyTorch official documentation~\cite{EffectiveTensorflow2, PyTorchTutorials}.}                                                                                            \\ \hline
\rev{D  }                                   &   \rev{Number of defects  }                           & \rev{Calculated via Pylint by summing all flagged code issues by severity (e.g., convention, warning, error).      }                                 \\ \hline
\rev{ADC}                                    & 	\rev{Annotation and documentation consistency}
& \rev{Calculating the ratio of elements (methods/classes) with type annotations and docstrings to the total number of defined functions and classes via AST. }                                                                                                                  \\ \hline
\rev{CH}                                    & \rev{Degree of method-field relation within a class} & \rev{Analyzed via field-method usage overlap using custom AST analysis.}                                                                                                             \\ \hline
\rev{CP }                                   & \rev{Degree of intra- and inter-module dependency.}           &  \rev{Calculated based on the number of inner and inter-module references/imports via AST.    }                                                                                                          \\ \hline
\rev{TC    }                                & \rev{Type inconsistencies between the usage and declaration of variables/functions}      & 
\rev{Calculated using mypy library~\cite{mypy_2024} to detect mismatches between declared type annotations and actual variable/function usage per PEP 484~\cite{PEP484}. }                                  \\ \hline
\rev{DQ }                                   &\rev{Proportion of entities with docstrings}          & \rev{Computed as the ratio of documented entities (methods/classes), using AST analysis.}                                                                                                                  \\ \hline
\rev{REP }                                  & \rev{Assesses a project's ability to produce consistent results by checking for key elements that support repeatable execution.  }   
&\rev{Computed based on the presence of a dependency file (requirements.txt), a README.md for instructions, and whether random initialization is explicitly controlled. Randomness is measured by checking the use of standard seeding functions (e.g., np.random.seed) and common initialization patterns across Python files in the project.}                                           \\ \hline
\end{tabular}
\caption{\rev{Low-level metrics used in PyQu: what they measure, how they are computed, and their mapping to high-level software quality attributes (QAs) defined in Sections 2.3.1–2.3.5. \textbf{Met}: Metric}}
\vspace{-0.8cm}
\label{tab:pyqu-metrics}
\end{table*}}

\subsubsection{\textbf{Reliability (RE)}} \rev{refers to the probability that a software system will operate without failure under specified conditions for a given period of time~\cite{1634994}. We assess reliability using a combination of low-level metrics commonly adopted in prior research, providing a solid basis for its automated evaluation ~\cite{ebert2016cyclomatic,quyoum2010improving,sahu2018revisiting,garnaik2014reliability,wen2019}. First, CC captures control-flow intricacy—higher values indicate more complex logic that is harder to test and more prone to errors, thereby reducing reliability~\cite{ebert2016cyclomatic}. Second, Type Checking (TC) also serves as an indicator of reliability, by detecting type-related inconsistencies at compile time; fewer violations suggest stronger type safety and more predictable runtime behavior~\cite{quyoum2010improving}. Third, the Defects (D) metric, derived from static analysis warnings and errors, quantifies deviations from coding standards—lower defect counts indicate a lower likelihood of runtime failures~\cite{sahu2018revisiting}. 
Fourth, SCS promotes code uniformity and reduces ambiguity, thereby improving reliability through consistent developer expectations~\cite{garnaik2014reliability}. Finally, Annotation and Documentation Consistency (ADC) indicates how well code comments and annotations match the actual implementation, helping reduce misunderstandings and errors during maintenance~\cite{wen2019}.}

We model (RE) as a function of the changes ($\Delta$) in the low-level metrics described, as defined in \autoref{eq:r_equation}.

\vspace{-1em}
\begin{equation}
    \text{RE} = \text{MLC}\big(\Delta CC,\Delta TC, \Delta D, \Delta SCS, \Delta ADC \big)
    \label{eq:r_equation}
\end{equation}

\subsubsection{\textbf{Maintainability (MA)}} \rev{refers to how easily a software system can be modified to correct defects, improve performance, or adapt to changes in the environment~\cite{610.12-1990}. We assess maintainability using low-level metrics that capture structural properties influencing the effort and risk associated with code changes. CC indicates the complexity of a program's logic, helping assess how difficult it is to understand and safely modify the code~\cite{ebert2016cyclomatic}. Lines of Code (LoC) provides a basic measure of system size—larger codebases are generally more difficult to maintain due to increased surface area for bugs and changes~\cite{bhatia2014loc}.  Cohesion (CH) reflects how strongly the responsibilities of a module are related; higher cohesion tends to improve maintainability by localizing functionality and reducing the mental load on developers~\cite{shih2001cohesion, ivers2024mind}. Coupling (CP) captures the degree of interdependence both between modules (external coupling) and within modules (internal coupling); lower coupling is desirable, as it reduces the ripple effect of changes and simplifies isolated updates~\cite{athanasopoulos2015coupling, ivers2024mind}. We model (MA) as a function of the changes ($\Delta$) in the low-level metrics described, as defined in \autoref{eq:m_equation}.}

\vspace{-0.9em}
\begin{equation}
    \text{MA} = \text{MLC}\big(\Delta CC, \Delta LoC, \Delta CH, \Delta CP \big)
    \label{eq:m_equation}
\end{equation}
\vspace{-2.5em}
\subsubsection{\textbf{Usability (US)}} \rev{reflects how easily developers can interact with, understand, and reuse a system—an essential attribute in collaborative ML workflows. We assess usability through three key factors: Documentation Quality (DQ), understandability (UN) metrics, and reproducibility (REP) indicators. Higher-quality documentation, such as the presence of meaningful docstrings, facilitates the effective reuse and execution of system components (e.g., modules and classes)~\cite{plösch2014documentation, treude2020documentation, tang2023documentation}. Code that is more understandable, as indicated by higher UN metrics scores, tends to be easier to navigate and adopt. Finally, REP is assessed via the presence of elements that enable repeatable execution—dependency files, documentation, and controlled randomness—where higher values indicate greater reproducibility and more consistent behavior across environments. Usability (US) is modeled as a function of the corresponding metric deltas ($\Delta$), as defined in Equation~\ref{eq:use_equation}.}

\vspace{-0.9em}
\begin{equation}
    \text{US} = \text{MLC}\big(\Delta UN, \Delta DQ, \Delta REP \big)
    \label{eq:use_equation}
\end{equation}
\subsubsection{\textbf{Modularity (MO)}}
\rev{refers to the degree to which a software system is decomposed into independent, well-structured components that enhance maintainability, and scalability~\cite{parnas1972criteria}. We assess modularity using low-level metrics that capture the separation of concerns and degree of interdependence among components. These include CH, which evaluates how closely related the responsibilities of a single module are~\cite{chidamber1994metrics}, higher cohesion suggests that a module is well-focused on a single responsibility, which improves maintainability and reduces unintended side effects. CC, which reflects structural simplicity and indirectly influences modular clarity~\cite{stevens1974structured}; and Coupling (CP), which we explicitly divide into \textit{internal} ($CP_{internal}$) and \textit{external} coupling ($CP_{external}$). $CP_{internal}$ measures interdependencies within a module or file, while $CP_{external}$ captures dependencies between different modules or external components~\cite{peggy2018exploring}. This separation aligns with established modularity principles in software engineering, where minimizing both internal entanglements and external dependencies is key to maintainability and scalability~\cite{chidamber1994metrics}.  We model (MO) as a function of the changes ($\Delta$) in the low-level metrics described, as defined in \autoref{eq:mod_equation}.}
\vspace{-0.75em}
\begin{equation}
    \label{eq:mod_equation}
    \text{MO} = \text{MLC}\big(\Delta CC, \Delta CH, \Delta CP_{\text{external}}, \Delta CP_{\text{internal}} \big)
\end{equation}

\vspace{-1.2em}
\subsection{PyQu Implementation \& Evaluation}

\subsubsection{\textbf{Extract Low-Level Metrics:}}
\rev{To ensure robust metric extraction, we used static analysis tools such as Radon~\cite{radon_2023}, Pylint~\cite{pylint_2024}, and others~(\autoref{tab:pyqu-metrics}) to measure control-flow complexity, style violations, type inconsistencies, and related attributes. Moreover, given the unique characteristics of ML codebases, we extended Pylint with custom checkers for ML-specific constructs in TensorFlow and PyTorch, aligning with their official best practices~\cite{EffectiveTensorflow2, PyTorchTutorials}. We also validated conformance with key Python Enhancement Proposals (PEP 8~\cite{PEP8}, PEP 257~\cite{PEP257}, PEP 484~\cite{PEP484}, and PEP 498~\cite{PEP498}). Metrics definitions and their computation methods are summarized in~\autoref{tab:pyqu-metrics}.}



\vspace{-0.5em}
\subsubsection{\textbf{Classifying Commits Using ML Models}:}
\label{subsub:classifycommits}

Once \textit{\toolName} calculates the low-level metrics, we classify whether a commit's changes improve the studied QAs using ML classification models rather than a weighted heuristic. Unlike ML models, weighted heuristics rely on subjective, manually assigned weights and fail to capture complex metric relationships, making them less adaptable to varying code changes~\cite{rivera2015incorporating}. \rev{Instead of manually balancing conflicting metric signals—such as when some low-level metrics improve while others decline—we rely on ML classifiers to learn such patterns automatically and learn non-trivial relationships between low-level metric changes and actual quality enhancement.}

Specifically, we employ three types of ML classification models—linear, non-linear, and ensemble learning models—previously used in related research~\cite{noei2024detecting}. These models include Adaptive Boosting (AB), CatBoost (CB), Decision Tree (DT), Gradient Boosting Machine (GBM), LightGBM (LGBM), Multi-layer Perceptron (MLP), Random Forest (RF), Support Vector Machine (SVM), and XGBoost (XGB). \rev{For each QA, each classifier uses as input features the full set of deltas—defined as the difference between post- and pre-code change values of each QA-related low-level metrics— enabling it to learn patterns in the data and identify which commit changes are most predictive of QAs improvements.}
Due to the uniqueness of the input features for each QA, we trained and tested the classifiers independently for each of the five QAs using a manually labeled dataset discussed in subsection \ref{subsub:dataset}


\rev{We tuned the hyperparameters for each (ML classifier–QA) pair and selected the best-performing classfier for each QA based on the evaluation metrics described in~\autoref{subsubsec:evaluation}. The full set of tested hyperparameters and the optimal settings for each ML classifier are available in our replication package~\cite{MLCodeQuality}.}
\vspace{-1.5em}
\subsubsection{\textbf{Training \& Testing Dataset}}
\label{subsub:dataset}
To construct the dataset used to train and test the ML classifiers used in \toolName, we employed Simple Random Sampling (SRS)\cite{baltes2022sampling}. 
Cochran’s formula \cite{Cochran1977} required 385 commits for a 95\% confidence level with a 5\% margin of error. To ensure robustness and generalizability in our evaluation, we increased our sample size to 500 randomly selected commits.

Two authors with expertise in software evolution and maintenance independently reviewed the commits, categorizing each as “Enhanced” or “Not Enhanced” against the five predefined \rev{QAs}.
\rev{To ensure consistency, the authors agreed that any commit exhibiting simultaneous evidence of both quality enhancement and quality degradation would be discarded; however, no such cases were found in the dataset.}
The annotation process began with each author independently examining 25\% of the sample—exceeding the 10\% minimum recommended threshold in empirical software engineering studies \cite{Campbell_Quincy_Osserman_Pedersen_2013}. A calibration meeting was held to discuss ambiguous cases and refine shared understanding. 
After reviewing all 500 commits, the authors reached a Cohen’s Kappa score of 0.87, indicating a high level of agreement in commit labeling \cite{Campbell_Quincy_Osserman_Pedersen_2013}.
A final adjudication meeting was conducted to resolve any remaining disagreements, ensuring a consensus-driven ground truth.

After the manual labeling process, we observed a notable data imbalance. For instance, in the context of understandability, only 133 out of 500 commits were labeled “Enhanced”. 
Due to the limited dataset size, downsampling would have exacerbated this imbalance. Instead, we applied the Synthetic Minority Oversampling Technique (SMOTE) \cite{Pradipta2021SMOTE}—a widely recognized method for handling imbalanced classification—that creates synthetic samples by interpolating between existing minority-class instances.
\rev{This approach enabled our classifiers to more effectively identify quality-enhancing changes without overfitting to the dominant class, aligning with prior findings that have demonstrated SMOTE’s efficacy and practicality in similar software engineering contexts \cite{Pradipta2021SMOTE,blagus2013smote, fernandez2018smote}}


As a result, we obtained the final balanced commit counts for each quality, with an equal split between “Enhanced” and “Not Enhanced”: Understandability (734 commits: 367 per category), Reliability (804 commits: 402 each), Maintainability (872 commits: 436 each), Usability (754 commits: 377 each), and Modularity (944 commits: 472 each). Subsequently, we split our corpus into training and testing sets, using 70\% of the dataset for training the classification models and 30\% to evaluate their performance on unseen data. 
\vspace{-0.8em}

\subsubsection{\textbf{PyQu Evaluation}}
\label{subsubsec:evaluation}
To ensure we selected the best classifiers against the studied \rev{QAs}, we carefully evaluated each of them to prevent overfitting and improve generalization. We used key performance metrics—accuracy, precision, recall, F1 score, as well as Receiver Operating Characteristic - Area Under the Curve (ROC-AUC) \cite{fawcett2006introduction}, and $\Delta$ Accuracy which is the absolute difference between training and test accuracy \cite{Yacouby2020Probabilistic}. To find the best settings for each model, we ran extensive hyperparameter searches and documented all configurations for transparency. We monitored overfitting by comparing training and testing accuracy \cite{Hoffer2017Train}, focusing on models with the smallest $\Delta$ accuracy rather than just the highest accuracy, ensuring reliable and robust results.
This quantitative assessment provided deeper insights into the PyQu’s effectiveness, highlighting its strengths and limitations in detecting quality enhancements within MLS projects.




\vspace{-1.1em}
\subsection{Quality-Enhancing Changes Identification}
\label{subsec:taxonomy}

After selecting the best classifiers and their optimized hyperparameters for each \rev{QA}, we executed PyQu against our dataset of \datasetSize commits.\ Next, we aimed to identify quality-enhancing changes in Python MLS, and map them with the \rev{QAs} they enhance. 
To accomplish this, we separately collected all commits labeled as enhancing a specific QA, forming five distinct sets. We then applied SRS with a 95\% confidence level and a 5\% margin of error to each set, yielding a representative subset of 1,728 commits: Understandability (350/3,863), Reliability (351/3,999), Maintainability (341/3,025), Usability (357/4,915), and Modularity (329/2,247). The percentage of duplicated commits across the five sample sets is low (18.37\%), ensuring an unbiased assessment of which change improves a specific QA or whether it impacts multiple QAs simultaneously.

\rev{To explore how specific changes drive quality improvements, we manually analyzed a representative subset of commits. Since many commits are non-atomic and include multiple edits, attributing improvements to individual changes is challenging for automated methods~\cite{behnamghader2017better}. Following the approach of Behnamghader et al.~\cite{behnamghader2017better}, we used manual review to identify the modifications most likely responsible for the detected enhancements. For example, commit \texttt{c11b191}~\cite{c11b191} included several updates to a CNN model, but our analysis identified one specific change—adding a descriptive name to a DNN layer—as the key contributor to improved understandability. We highlight such changes not as sole causes, but as recurring patterns consistently observed through manual inspection.
}

The labeling was based on an open-coding process that allowed code changes to emerge from the data rather than preconceptions.
First, during the planning phase, two authors agreed to manually review the subset of commits to determine the change types and verify whether they truly enhance the corresponding QAs as labeled by PyQu.
Second, for the Identification round, the two co-authors separately observed all the commit changes to identify any code change. They used low-level metrics, commit messages, titles, and descriptions to verify PyQu labeling and confirm the changes impacted the labeled QAs. The labelers kept a shared record of high-level descriptions of the changes serving as a dynamic reference, containing for each observed change a textual definition, a candidate name, and an illustrative example to reduce bias/information leakage between the labelers and avoid any redundant definitions.
For the categorization round, the labelers categorized the code changes identified in the first round following the recommendations outlined by Usman et al. concerning the construction of taxonomies \cite{usman2017taxonomies}. After the two labeling rounds concluded, Cohen’s Kappa score reflecting the agreement on the identification and the classification of the quality-enhancing code changes was calculated at 0.81, signaling high agreement \cite{Campbell_Quincy_Osserman_Pedersen_2013}. Then, three rounds of consensus meetings were carried out to resolve any disagreements. Third, for the design and construction phase, a card-sorting process~\cite{fincher2005making,spencer2004card, ghammam2025build} was performed by the two authors to group changes of similar characteristics into \changes\ types categorized into \categories\ categories for easier generalizability and usability.
Finally, for the Validation phase, a third co-author with extensive expertise in software quality and maintenance, confirmed the applicability of the various categories, ensuring they reflect significant recurring patterns and constitute valuable knowledge for developers.

\section{Study Results}
\label{sec:Results}
\subsection{PyQu Effectiveness (RQ1)}
\label{subsec:RQ1}

To address RQ1, we applied the methodology outlined in subsections \ref{subsub:classifycommits}, \ref{subsub:dataset}, and \ref{subsubsec:evaluation}.  
The results, presented in \autoref{tab:performance}, illustrate the accuracy (Acc.), precision (Prec.), recall (Rec.), F1-score (F1), ROC AUC (RC), and train-test accuracy difference ($\Delta$ Acc.) scores of the best-performing models across each of the studied \rev{QAs}, using their optimal hyperparameters. The models were selected based on their superior accuracy and stability. The hyperparameters and results of all models can be found in \cite{MLCodeQuality}.

PyQu demonstrates strong effectiveness in detecting quality-enhancement changes in MLS code, achieving consistently high accuracy across the five \rev{QAs}. The accuracy ranges from 0.82 for Understandability and Usability with Random Forest and CatBoost respectively, to 0.84 and 0.86 for Reliability and Modularity both using CatBoost, and 0.87 for Maintainability with LightGBM, highlighting its reliability across the five studied quality aspects.





Furthermore, PyQu achieved both high precision and recall across the five \rev{QAs}. The high precision scores (0.82–0.88) indicate that the model effectively minimizes false positives, ensuring that identified quality-enhancing commits are genuinely impactful. Meanwhile, high recall scores (0.80–0.89) demonstrate PyQu effectiveness in minimizing false negatives, reflecting its ability to capture most quality-enhancing commits.



Overall, these results validate that PyQu is highly effective in detecting quality-enhancing modifications in ML projects,while minimizing bias and misclassification errors.
\vspace{-1em}
\begin{table}[ht!]
\centering
\begin{tabular}{|c|c|c|c|c|c|c|c|}
\hline
\textbf{Quality}     & \textbf{Model}        & \textbf{Acc.}       & \textbf{Prec.}      & \textbf{Rec.}       & \textbf{F1}       & \textbf{RC}  &  $\Delta$\textbf{ Acc.} \\ \hline
UN          & RF                      & 0.82	          & 0.85	            & 0.80	             & 0.82	             & 0.86	             & 0.05 \\ \hline
RE         & CB                      & 0.84	          & 0.86	            & 0.84	             & 0.85	             & 0.88              & 0.02 \\ \hline
MA           & LGBM                    & 0.87	          & 0.86	            & 0.89	             & 0.87              & 0.91	             & 0.05 \\ \hline
US           & CB                      & 0.82	          & 0.82	            & 0.85	             & 0.83              & 0.84	             & 0.01 \\ \hline
MO          & CB                      & 0.86	          & 0.86	            & 0.88               & 0.87	             & 0.91	             & 0.02 \\ \hline
\end{tabular}
\caption{Performance metrics for the studied Qualities. }
\label{tab:performance}
\vspace{-2.5em}
\end{table}

\vspace{-1em}
\subsection{Model Generalizability (RQ2)}
\label{subsec:RQ2}

The results in \autoref{tab:performance} clearly demonstrate that PyQu adapts effectively to unseen code changes without overfitting. The small training–test accuracy gap ($\Delta$ Acc.), ranging from 0.01 to 0.05 across quality attributes, remains below the common 0.1 threshold in ML \cite{allamanis2018survey, hellendoorn2017deep}. This indicates that the model is effectively learning meaningful patterns rather than overfitting to the training data. For instance,$\Delta$ Acc. stands at 0.05 for Understandability and Maintainability, 0.02 for Reliability and Modularity and reaches 0.01 for Usability. These results suggest that PyQu captures the real mechanics of code quality enhancements instead of merely memorizing examples, which reinforces the robustness of its predictions.

Beyond accuracy, the tool attains a strong balance between precision and recall across different quality attributes. For instance, modularity achieves a precision of 0.86 and recall of 0.88, while usability maintains a precision of 0.82 and recall of 0.85. This balance translates into high F1 scores ranging from 0.82 to 0.88. This balance is critical for ensuring reliable classification across all quality attributes, reinforcing PyQu’s trustworthiness for real-world deployment.

The ROC AUC scores further validate PyQu's generalizability. Maintainability and Modularity achieved the highest ROC AUC scores (0.91), followed by Reliability (0.88), Understandability (0.86), and Usability (0.84). These results indicate strong cross-validation stability, confirming that \toolName~maintains high discriminative power across diverse input distributions, which ensures reliable and consistent performance in detecting quality-enhancing code changes.


Overall, these results show that PyQu can generalize well, which makes it a reliable tool for finding MLS quality improvements, even when looking at code changes that haven't been seen before.

\subsection{Quality-Enhancing Changes in Python MLS (RQ3)}
\label{subsec:RQ3}


To identify and categorize the quality-enhancing code changes in Python-based MLS, we adopted the \rev{manual labeling} methodology presented in \autoref{subsec:taxonomy}. Our analysis revealed that 16.4\% of the commits were false positives, often only adding new functionality or adjusting logic.
In total, we identified 2,338 code changes spanning \changes\ unique types, encompassing both already established and novel modifications. These code changes were then organized into \categories\ categories, each reflecting the broader context in which they occur. \autoref{tab:change} illustrates detailed information about these changes, including their categories, observed frequencies (F), and associated quality attributes improved by each change. Furthermore, the table indicates whether each change is detected by PyRef \cite{pyref}, Python RefMiner \cite{PyRefMiner}, and R-CPatMiner \cite{Dilhara_Ketkar_Sannidhi_Dig_2022}, the state-of-the-art Python changes detection tools.
In the subsequent section, we detail these code changes by category, highlighting key novel modifications. A full list of the identified code changes is available in our replication package \cite{MLCodeQuality}.


\begin{table*}[ht]
    \centering
    \small
    \renewcommand{\arraystretch}{0.9}
    \caption{Quality-enhancing code changes taxonomy. Column F shows the frequency of the discovered changes. Column N indicates newly discovered changes. Columns PR, RM and CP indicate if the change was identified by PyRef, Python RefMiner, and R-CPatMiner respectively. And the last 5 columns represent the quality attributes}
    \vspace{-0.3cm}
    \begin{tabular}{|c|l|c||c|c|c|c||c|c|c|c|c|}
        \hline
         \rowcolor{gray!40} \textbf{Category} & \textbf{Change Name} &  \textbf{F} & \textbf{N} & \textbf{PR} & \textbf{RM} & \textbf{CP} & \textbf{UN} & \textbf{RE} & \textbf{MA} & \textbf{US} & \textbf{MO} \\
        \hline

        \multirow[c]{3}{*}{Import Restructure}  &  Replace Local Imports with Global Scope   & 3 &\ding{51} & &   & &  & \ding{51} &  & \ding{51} & \\  \cline{2-12}
                                                & Replace Relative Imports with Explicit Imports & 14 &\ding{51} & &   & & \ding{51} & \ding{51} &  \ding{51} &    & \ding{51}\\  \cline{2-12}
                                                &  Replace Explicit Imports with Relative Imports & 5 & \ding{51} & &  &  &  & \ding{51} & \ding{51} & \ding{51} &   \\ 
        \hline
        \multirow[c]{5}{*}{Code Simplification}&  Loop Index Simplification     & 16 & &   &  & & \ding{51} & \ding{51} &  & \ding{51} &\\ \cline{2-12}
                                                                                & Replace Repeated Code with Dynamic Execution  & 9 &\ding{51} & &   & & \ding{51} & \ding{51} & \ding{51} & \ding{51} &\\ \cline{2-12}
                                                                                & Condition Simplification    & 34 &\ding{51}  & &   & \ding{51}& \ding{51} & \ding{51} & \ding{51} & \ding{51} &  \\ \cline{2-12}
                                                                                & Replace Attribute Check with Explicit Type Check     & 3 &\ding{51} & &   & &  & \ding{51} &  &  & \\ \cline{2-12}
                                                                                & Replace Equality Check with Identity Check   & 2 &\ding{51} & &   & &  & \ding{51} &  & & \\
        \hline
        Migration to Supported API              & Replace Deprecated API Usage   &6 & \ding{51} & &    & & \ding{51} & \ding{51} & \ding{51} & \ding{51} &  \\ 
        \hline
        \multirow[c]{4}{*}{Use Optimized Features}   & Use Optimized Built-in Method  & 28 & & &   & \ding{51} & \ding{51} & \ding{51} & \ding{51} & \ding{51} & \\ \cline{2-12}
                                                    &  Use Optimized Library Method  &29 &   & &   &\ding{51} & \ding{51} & \ding{51} & \ding{51} & \ding{51} & \\ \cline{2-12}
                                                    & Replace Iterable with Container  & 7 & & &   &\ding{51} &  & \ding{51} &  &  & \\ 
        \hline
        \multirow[c]{4}{*}{Enhance Documentation}   & Add Type Annotation  & 64 & \ding{51} & &   & & \ding{51} & \ding{51} & \ding{51} & \ding{51} & \\ \cline{2-12}
                                                    &  Document Code  & 214 & & &    & & \ding{51} & \ding{51} &  & \ding{51} & \\ \cline{2-12}
                                                    & Enhance Argument Parsing Description  & 4 & \ding{51} & &   & & \ding{51} &  &  & \ding{51} & \\ \cline{2-12}
                                                    & Add DNN Layer Name  & 2 & \ding{51} & &   & & \ding{51} &  &  & \ding{51} & \\
        \hline
        \multirow[c]{3}{*}{String Formatting}       &  Replace \% with .format()   &3 & \ding{51} & &   & &  &  & \ding{51} &  &\\ \cline{2-12}
                                                    & Replace .format() with f-strings &1 & \ding{51} & &   & & \ding{51} &  &  &    & \\ \cline{2-12}
                                                    &  Introduce f-string &5 &\ding{51} & &   & & \ding{51} &  &  &  &   \\ 
        \hline
        \multirow[c]{3}{*}{Enhance Error Handling } & Introduce Exception Handling   & 6 & & & \ding{51}   & &  & \ding{51} &  &  & \\ \cline{2-12}
                                                    & Replace Generic Exception with Specific Exception & 1 & \ding{51} & &   & & \ding{51} &  &  &   & \\ \cline{2-12}
                                                    & Replace Specific Exception with Generic Exception & 1 &\ding{51} & &   &  & & \ding{51} &  &  &   \\ 
        \hline
        \multirow[c]{5}{*}{Data Reorganization}     & Replace Hardcoded Value with Parameter  & 6 & & \ding{51} &  \ding{51} & & \ding{51}  &  & \ding{51} & \ding{51} &  \\ \cline{2-12}
                                                    & Replace Hardcoded Value with Variable  & 22 & & &    & & \ding{51} &  & \ding{51} & \ding{51} & \\ \cline{2-12}
                                                    & Replace Hardcoded Value with Constant  & 8 & & &   & & \ding{51} &    & \ding{51} & \ding{51} &\\ \cline{2-12}
                                                    & Replace Variable with Hardcoded Value  & 8 & & &    & & \ding{51} &  &  &  &  \\ \cline{2-12}
                                                    & Replace Parameter with Attribute   & 4 & & &   & &  &  & \ding{51} &  & \ding{51} \\ \cline{2-12}
                                                    & Replace Attribute with Parameter  & 2 & & & \ding{51}  &  & &  & \ding{51} &  & \ding{51}\\ 
        \hline
            \multirow[c]{5}{*}{Function Signature Refinements}  & Introduce Parameter Default Value         & 12 &\ding{51} & &   & & \ding{51} & \ding{51} &           & \ding{51} & \\ \cline{2-12}
                                                                & Remove Parameter Default Value            & 6  &\ding{51} & &   & &           &           & \ding{51} & \ding{51} & \\ \cline{2-12}
                                                                & Explicitly Naming Function Parameters     &7 &\ding{51}  & &   & & \ding{51} & \ding{51} &  & \ding{51} & \\ \cline{2-12}
                                                                & Replace kwargs with Explicit Parameters   & 11 &\ding{51} & &   & & \ding{51} & \ding{51} & \ding{51} & \ding{51} & \\ \cline{2-12}
                                                                & Replace Explicit Parameters with kwargs   & 5 &\ding{51} & &   & &  & \ding{51} &  &  &  \\
        \hline

        \multirow[c]{3}{*}{Change Scope}    & Change Name or Variable Scope                 &2 & & &   &\ding{51} &  & \ding{51} &  &  & \\ \cline{2-12}
                                            & Replace Local Variable with Attribute         & 3 & & &   &  & & & \ding{51}  &  & \ding{51}\\ \cline{2-12}
                                            & Replace Static Method with Instance Method    & 2 &\ding{51}  & &   &  & &  &  &  &\ding{51} \\ 
        
        \hline 
        \multirow[c]{15}{*}{\shortstack[c]{Structure \\ Reorganization}}    & Extract Class & 82 & &  &  \ding{51}  & &  \ding{51} & \ding{51} & \ding{51} & \ding{51}  & \ding{51}\\ \cline{2-12}
                                                                            & Move Class & 8  & & &  \ding{51} & &   \ding{51} &  &  \ding{51} & \ding{51} &  \ding{51} \\ \cline{2-12}
                                                                            & Inherit Superclass   & 7 & & &  & &  \ding{51}  & \ding{51} &  \ding{51} &  \ding{51} &  \ding{51}\\ \cline{2-12}
                                                                            & Extract Superclass   & 37 & & &  \ding{51} & &  \ding{51} & \ding{51} & \ding{51}  & \ding{51} &  \ding{51} \\ \cline{2-12}
                                                                            & Extract Variable & 39 & &   &\ding{51} & & \ding{51} & \ding{51} & \ding{51} & \ding{51}   & \ding{51}\\ \cline{2-12}
                                                                            & Extract Method & 289 & & \ding{51} & \ding{51}  & &  \ding{51} & \ding{51} & \ding{51} &  \ding{51}  & \ding{51}\\ \cline{2-12}
                                                                            & Move Method & 50 & & \ding{51}&  \ding{51} &  & \ding{51} & \ding{51} & \ding{51} & \ding{51} & \ding{51}  \\ \cline{2-12}
                                                                            & Inline Method & 60 & & \ding{51}& \ding{51}  & & \ding{51} & \ding{51} & \ding{51} & \ding{51}  & \ding{51}\\ \cline{2-12}
                                                                            & Externalize Model Configuration   & 4 &\ding{51} & &   & &   &  & \ding{51} &  & \ding{51}\\ \cline{2-12}
                                                                            & Push Down Method & 2 &  & \ding{51} & \ding{51}  & &  &    & \ding{51} &  & \ding{51}\\ \cline{2-12}
                                                                            & Extract And Move Method & 14  &  & &  \ding{51} & & \ding{51} & \ding{51} & \ding{51} &\ding{51}  & \ding{51}\\ \cline{2-12}
                                                                            & Move Variable & 3 & & &   &  & \ding{51} &  &  \ding{51}  & \ding{51} & \ding{51}\\ \cline{2-12}
                                                                            & Extract Module & 51 & & &   & & \ding{51} & \ding{51} &  \ding{51}  & \ding{51} &\ding{51} \\ \cline{2-12}
                                                                            & Extract And Pull Up Method & 4 & &  &    & &  &  & \ding{51} &  & \ding{51}\\ \cline{2-12}
                                                                            & Pull Up Method & 2 & & \ding{51} &  \ding{51} & &  &    & \ding{51} &  & \ding{51}\\ \cline{2-12}
        \hline

        \multirow[c]{6}{*}{Clean up}            & Remove Unused Imports & 132 & \ding{51} & &   &  &\ding{51} & \ding{51} & \ding{51} & \ding{51} & \ding{51}\\ \cline{2-12}
                                                & Remove Variable   & 28 & & &  \ding{51} & & \ding{51} &  & \ding{51} & \ding{51} & \\ \cline{2-12}
                                                & Remove Parameter & 56 & & \ding{51}  & \ding{51} & & \ding{51} &  & \ding{51} & \ding{51} &   \\ \cline{2-12}
                                                & Remove Redundant Parentheses & 4 & \ding{51} &  &    & &  \ding{51}  &  & \ding{51} &  \ding{51} & \\ \cline{2-12}

                                                & Remove Dead Code & 217 &  & &   & & \ding{51} & \ding{51} & \ding{51} &  \ding{51}  & \\ \cline{2-12}
                                                & Reformat Code & 184 &  & &   & & \ding{51}  &  & \ding{51} & \ding{51} & \\ \cline{2-12}
        \hline

        \multirow[c]{3}{*}{Rename Identifier}   & Rename Class   & 29 & & &  \ding{51} & & \ding{51} &  & & \ding{51} & \\ \cline{2-12}
                                                & Rename Method   & 194 & &  \ding{51} & \ding{51} & & \ding{51} &  &  & \ding{51} & \\ \cline{2-12}
                                                & Rename Variable   & 205 & & &  \ding{51} & & \ding{51} &  &  & \ding{51} & \\ \cline{2-12}
                                                & Rename Parameter   & 76 & & \ding{51} & \ding{51}  & & \ding{51} &  & & \ding{51} & \\
        \hline


    \end{tabular}
    
    \vspace{-0.2cm}
    \label{tab:change}
\end{table*}


\subsubsection{\textbf{Imports Restructure}} We identified 22 instances in this category, encompassing a set of novel changes aimed at improving how libraries are managed throughout the codebase. \textit{`Replace Local Imports with Global Scope'}, accounts for 13.6\% (3/22) of the instances in this category. This change involves moving imports that were originally placed inside methods or classes to the top of the file. By centralizing dependency management, this approach aligns with best practices for importing libraries as outlined in PEP 8 \cite{PEP8}. It enhances usability and reliability by providing a clear and consolidated view of all dependencies in use and making sure it's imported only once.

\autoref{lst:RelativeImport} demonstrates another change within this category: \textit{`Replace Relative Imports with Explicit Imports'} (14/22). In this example, the module \textit{nn} (line 1) from the Torch library was only used once to call \textit{CrossEntropyLoss()} (line 3). Instead of importing the entire module, an explicit import (line 2) was introduced, specifying only the required method. By including only used modules and methods, this refinement removes ambiguity, makes dependencies clearer, and improves code understandability, maintainability, reliability, and modularity. 

On the other hand, \textit{`Replace Explicit Imports with Relative Imports'} represents 5 instances (22.7\%). It converts explicit imports into relative ones, a practice particularly useful in projects with complex package structures. Switching to relative imports enhances reliability by ensuring the inclusion of all modules, especially in complex systems, consequently making the code more maintainable and usable. 

 \vspace{-0.7em}

\begin{center} 
\begin{lstlisting}[language=diff,caption={Commit \textit{b35c670a} in kenshohara/3D-ResNets-PyTorch: Replace Relative Imports with Explicit Imports.},label={lst:RelativeImport},numbers=left,frame=lines, escapechar=!, ,columns=fixed]
!\RedL!from torch import nn
!\GreenL!from torch.nn import CrossEntropyLoss
!\RedL!criterion = nn.CrossEntropyLoss().to(opt.device)
!\GreenL!criterion = CrossEntropyLoss().to(opt.device)
\end{lstlisting}
\vspace{-0.5em}
\end{center}

\subsubsection{\textbf{Code Simplification}}
This category, accounting for 64 instances, focuses on streamlining existing code constructs and logic flows to reinforce maintainability, understandability, and reliability. \textit{`Loop Index Simplification'} (representing 25\% (16/64)) employs built-in functions—such as Python’s \textit{enumerate()}—to eliminate manual index tracking and reduce off-by-one errors. In a similar vein, the newly identified change representing 14\% (9/64) of instances in this category, \textit{`Replace Repeated Code with Dynamic Execution'}, consolidates repetitive patterns (often via loops), cutting down on duplication and potential sources of error. For instance, as shown in \autoref{lst:duplicate}, the developer encapsulates the shared behavior of four tasks (as in line 4) with a single test function (lines 4-5) by looping on their names. This type of change makes the code easier to understand and maintain, while enhancing reliability and usability. Moreover, the newly discovered change, \textit{`Condition Simplification'}, accounts for 53\% (34/64) of the changes in this category. It simplifies complex `if statements', making control flows more understandable, usable, reliable, and easier to maintain. Additionally, a newly discussed change, \textit{`Replace Attribute Check with Explicit Type Check'} (3 instances), enhances reliability by clearly verifying data types, bypassing unexpected polymorphic behaviors. 
Similarly, \textit{`Replace Equality Checks with Identity Checks'} (2 instances) improves reliability by shifting from value-based comparisons to object reference comparisons. This change helps prevent unintended equality evaluations, reducing potential bugs and enhancing overall system reliability.


\begin{center} 
\vspace{-1em}
\begin{lstlisting}[language=diff,caption={Commit \textit{b01c4d935} in okfn-brasil/serenata-de-amor: Replace Repeated Code with Dynamic Execution.},label={lst:duplicate},numbers=left,frame=lines, escapechar=!, ,columns=fixed]
!\GreenL!SITUATIONS = [('ABERTA', False), ('BAIXADA', True), ...]
def test_is_regular_company(self):
!\RedL!  self.assertEqual(.. 'situation': 'ABERTA', False)
!\GreenL!  for situation in self.SITUATIONS:
!\GreenL!    self.assertEqual(.. 'situation': situation[0], situation[1])
\end{lstlisting}
\end{center}
\vspace{-1em}

\subsubsection{\textbf{Migration to Supported API}}

This category includes a single operation occurring in 6 instances: \textit{`Replace Deprecated API Usage'}. This newly discovered change updates outdated built-in and third-party library APIs with newer, more efficient alternative. Such updates ensure alignment with current best practices, prevent issues from deprecated functionality, and collectively improve understandability, usability, maintainability, and reliability. For example, in commit \textit{7f6808fb} \cite{7f6808f} from the hunkim/PyTorchZeroToAll repository, the author replaced loss.data[0] with loss.item(). In earlier versions of PyTorch, loss.data[0] was commonly used to extract scalar values from a tensor. However, this approach bypassed PyTorch’s computation graph, which could lead to issues with autograd and gradient tracking \cite{PyTorch}. The recommended loss.item() method, introduced in later versions, safely extracts the value while maintaining compatibility with PyTorch’s dynamic computation graph \cite{PyTorch}.

\subsubsection{\textbf{Use Optimized Features}}
This category, with 64 changes identified, refines performance and reliability by leveraging more efficient data structures, built-in functions, and specialized libraries. \textit{`Use Optimized Built-in Methods'} (28 instances), such as employing OS-agnostic path manipulations (\texttt{os.path.join()}), eliminates platform-specific slash inconsistencies, thus enhancing the reliability of the code. Similarly, \autoref{lst:OptFeat} demonstrates how \textit{`Use Optimized Library Method'} (29 instances) adopts targeted utilities—such as \texttt{np.mean()} (Line 6) a NumPy enhanced method—instead of manual loops for computing arithmetic means, resulting in more concise code and reducing opportunities for error. The use of optimized built-in methods and third-party library functions enhances code understandability and usability, improves reliability through proven optimizations, and ultimately increases maintainability.
Finally, \textit{`Replace Iterable with Container'} (7 instances) aligns data usage with data structures that minimize overhead and reinforce reliability.

\subsubsection{\textbf{Enhance Documentation}} (12.1\% of the total identified changes), involves refining code clarity, and interpretability, ultimately improving how developers interact with software artifacts. \textit{`Add Type Annotation'} (64 instances) is a newly identified change that clarifies function contracts and data expectations by adding type annotations for variables, parameters, and the return of a method. This operation enhances understandability, usability, reliability, and maintainability \cite{PEP483}. \textit{`Document code'} (214 instances), further eases knowledge transfer and future modifications, by adding and expanding docstrings, and comments, aligning with best practices that emphasize explicit explanations, thus enhancing understandability, reliability, and usability of the code \cite{venkatkrishna2023docgen}. Similarly, the newly identified change, \textit{`Enhance Argument Parsing Description'} (4 instances), improves system usability and understandability by clarifying command-line parameters or function calls, thereby reducing user errors and streamlining interactions. Finally, \textit{`Add Deep Neural Network (DNN) Layer Name'} (2 instances) is another novel change observed in commit \textit{c11b191} \cite{c11b191} from broadinstitute/keras-rcnn. In this example, the developer adds descriptive depth by assigning intuitive names to the DNN layers, further supporting understandability and usability by providing intuitive identifiers. 

\begin{center} 
\vspace{-0.5em}
\begin{lstlisting}[language=diff,caption={Commit \textit{0e621bb8} in tensorflow/models: Use Optimized Library Method.},label={lst:OptFeat},numbers=left,frame=lines, escapechar=!, ,columns=fixed]
!\RedL!for (t, p) in zip(truth, predictions):
!\RedL!   if t == p: correct_prediction += 1
!\RedL!   total_prediction += 1
!\RedL!precision = float(correct_prediction) / total_prediction
!\GreenL!precision = np.mean(truth == predictions)
\end{lstlisting} 
\vspace{-0.4em}
\end{center}

\subsubsection{\textbf{String Formatting}} This category consists of all newly discovered changes that focus on enhancing string construction methods to produce more manageable code. In total, 9 changes were identified in this category, 33.3\% of them are represented by \textit{`Replace \% with .format()'}. This operation substitutes traditional `\%' python formatting with more flexible `.format()' method, streamlining the process of inserting values into strings, thereby improving maintainability. The second change, \textit{`Replace .format() with f-strings'}, presenting 11\% of the total number of changes in this category, transitions the code to f-strings, which allow developers to embed expressions directly within string literals, reducing verbosity while enhancing understandability \cite{PEP498}. 
Finally, \textit{`Introduce f-strings'} (55\%) replaces simple print statements with comma-separated values or string concatenation with f-string formatting, optimizing code by enhancing readability and reducing errors when embedding variables. This improves string construction quality and overall code understandability \cite{PEP498}.
\vspace{-0.2em}
\subsubsection{\textbf{Enhance Error Handling}} This category focuses on capturing, clarifying, and resolving unexpected runtime issues to ensure robust software behavior. In total, we identified 8 instances of three change types in this category. \textit{`Introduce Exception Handling'} (6 instances) allows systems to gracefully recover from anomalous states, promoting reliability by preventing catastrophic failures. \textit{`Replace Generic Exception with Specific Exception'} is a novel change that appeared once in our study. One example is illustrated in commit \texttt{e018c816} \cite{e018c81} from scikit-bio/scikit-bio repository. The author replaced \textit{except} serving as a general exception handling with a Pandas specific error, \textit{`except pd.core.common.PandasError`}. This further improves understandability, as pinpointing the exact error type supports quicker troubleshooting and fosters a cleaner architecture. 
Conversely, \textit{`Replace Specific Exception with Generic Exception'}, a new change observed once, involves substituting specific exceptions with broader generic exceptions. This practice improves reliability, especially in scenarios where error types are uncertain or unknown, ensuring robust error handling even under unexpected conditions.

\subsubsection{\textbf{Data Reorganization}}
This category focuses on restructuring data usage patterns, with 50 instances identified. 
\textit{`Replace Hardcoded Value with Parameter'} (6 instances) and \textit{`Replace Hardcoded Value with Variable'} (22 instances) help eliminate "magic numbers" from the code, which simplifies updates while improving understandability, usability, and maintainability. Similarly, \textit{`Replace Hardcoded Value with Constant'} (8 instances) enhances semantic clarity by explicitly defining values, making the code more readable, and reducing the risk of unintended modifications. Conversely, \textit{`Replace Variable with Hardcoded Value'} (8 instances) can simplify certain cases and improve understandability when a hardcoded value suffices, thereby reducing unnecessary abstraction. Lastly, \textit{`Replace Parameter with Attribute'} and \textit{`Replace Attribute with Parameter'} (4 and 2 instances, respectively) are opposite operations. Still, both relocate data to the most appropriate context—either centralizing it as an attribute or abstracting it via a parameter. These changes reduce duplication, improving maintainability and modularity.



\subsubsection{\textbf{Function Signature Refinements}} This category, encompassing 5 newly identified changes (41 instances), aims to refine function declaration and utilization, ultimately enhancing both clarity and robustness in MLS. \textit{`Introduce Parameter Default Value'} (12 instances) allows developers to define default values for parameters, ensuring reasonable fallbacks when arguments are omitted. This promotes understandability, usability, and reliability. In contrast, \textit{`Remove Parameter Default Value'} (6 instances) forces explicit declarations, thereby reducing ambiguity and reinforcing usability and maintainability for scenarios demanding precise parameter inputs, trading off reliability. Moreover, \textit{`Explicitly Naming Function Parameters'} (7 instances) highlights each argument’s role by explicitly stating the parameter during method invocation, boosting the code to be more predictable and less prone to errors, especially when library updates might alter default behaviors. \textit{`Replace kwargs with Explicit Parameters'} (11 instances) is about replacing the generic unpacking structure `kwargs' with specific parameters. This change simplifies type inference and enriches the understandability, usability, maintainability, and reliability of the code. Conversely, \textit{`Replace Explicit Parameters with kwargs'} (5 instances) proves useful when the number of arguments varies between calls, enabling dynamic method invocation, ultimately enhancing the reliability of the system.
\vspace{-0.8em}
\subsubsection{\textbf{Change Scope}} We identified 7 instances of changes that aim at switching the scope of an identifier. \textit{`Change Name or Variable Scope'} (2 instances) is a common change that was highlighted by Dilhara et al. \cite{Dilhara_Ketkar_Sannidhi_Dig_2022}, where a `\textit{with}' context manager controls the life-cycle of a code block. This is illustrated by commit \texttt{c541ac34}\cite{c541ac34} from analysiscenter/batchflow repository, which demonstrates how \textit{tf.variable\_scope(name)} improves reliability by providing a more controlled execution context. \textit{`Replace Local Variable with Attribute'} (3 instances) enhances the modularity and maintainability by binding the variable directly to its class. Finally, \textit{`Replace Static Method with Instance Method'} (2 instances) shifts a static method to an instance-level call, further promoting modularity.

\vspace{-0.9em}
\subsubsection{\textbf{Structure Reorganization}}This category highlights various code restructuring efforts, where we identified 652 change instances.
\textit{`Externalize Model Configuration'} (4 instances) is a newly identified change shown in \autoref{lst:MotEx}. In this example, developers moved model configuration details to a separate file, making them easier to manage and modify. This change improves maintainability and code modularity by centrizling and simplifying future updates.

Beyond configuration management, developers applied a variety of well-known restructuring techniques. At the class level, changes like \textit{`Extract Class'} (82/652) helped group related functionality more intuitively, making it easier to understand, use, maintain, and more modular and reliable. Additionally, \textit{`Move Class'} (8/652) ensured that classes fit better within the system architecture, enhancing understandability, maintainability, usability, and modularity.


At the inheritance structures. \textit{`Inherit Superclass'}—where a class inherits from an existing class—(7 instances) and \textit{`Extract Superclass'}—where a new superclass is extracted and inherited—(37 instances) were employed to consolidate shared behaviors. By reducing duplication, they improve the modularity, usability, reliability, understandability, and maintainability of the class hierarchy.

At the method and variable levels, techniques including \textit{`Extract Variable'} (39 instances) and \textit{`Extract Method'} (289 instances), both of which simplify code by isolating reusable logic into self-contained units. \textit{`Move Method'} (50 instances), \textit{`Inline Method'} (30 instances), \textit{`Extract and Move Method'} (14 instances), and \textit{`Extract Module'} (51 instances) further enhanced code organization—improving understandability, usability, reliability, maintainability, and modularity. Additionally, \textit{`Push Down Method'} (2 instances), \textit{`Pull Up Method'} (2 instances), and \textit{`Extract And Pull Up Method'} (4 instances) contributed to better maintainability and modularity. Finally, \textit{`Move Variable'} (3 instances) improved modularity within class hierarchies, understandability, usability, and maintainability.

\vspace{-0.5em}
\subsubsection{\textbf{Code Clean Up}} This category ( presenting 21.3\% (132/621) of the total number of changes) covers operations removing redundant or unnecessary elements of the code as well as restructuring it for better presentation. \textit{`Remove Unused Imports'} (21.3\%) involves removing unused libraries, such as in the commit \textit{a5bda34f}\cite{a5bda34} in deepset-ai/FARM. In this example, the developer removed \textit{import pandas as pd}, which was never used in that file. This change makes the system more understandable, modular, and usable, as it reduces unnecessary dependencies. Furthermore, it enhances reliability since the module becomes lighter by eliminating redundant imports, reducing potential conflicts, and improving execution efficiency. \textit{`Remove Variable'} (28 instances) eliminates unnecessary variables, reducing cognitive load and making the codebase more concise. Similarly, \textit{`Remove Parameter'} (56 instances) ensures that function signatures accurately reflect their intended behavior, preventing misleading or overly complex interfaces. Additionally, \textit{`Remove Redundant Parentheses'} (4 instances) is a newly discussed code change that streamlines code by eliminating superfluous parentheses, such as those in class head declarations. All these changes help the code become more understandable, easier to maintain, and easier to use.
\textit{`Remove Dead Code'} —which accounts for 34.9\% (217/621) of the instances in its category—eliminates code that is no longer executed. This directly enhances understandability, usability, and maintainability while also improving system reliability by removing outdated logic paths that could lead to unexpected issues in the future. Finally, \textit{`Reformat code'} (representing 29.6\% (184/621))—such as breaking lengthy lines, adding spacing between operands, or fixing arguments indentation—significantly improves readability. Each of those seemingly small visual improvements hugely contributes to understandability and maintainability, allowing faster debugging and overall better usability by developers.
\vspace{-0.5em}
\subsubsection{\textbf{Rename Identifier}}
This category accounts for 21.5\% of total changes (504 instances) and aims at renaming the different identifiers. \textit{`Rename Class'} (29 instances) modifies class names to better reflect their functionality, enhancing navigation and system understandability \cite{PEP8}. In a like manner, \textit{`Rename Method'} (194 instances) modifies method names to correspondingly meaningful and expressive ones, hence improving the understandability and consistency of the codebase. \textit{`Rename Variable'} and \textit{`Rename Parameter'} (205 and 76 instances, respectively) help ensure that variables and parameters have descriptive and meaningful names, following best practices in eliminating vague naming \cite{PEP8}.

Overall, 14.8\%, 31\%, and 8.2\% of the changeS identified in our research were detected by PyRef, Python RefMiner, and R-CPatMiner as shown in \autoref{tab:change}, respectively. While this demonstrates PyQu's effectiveness in identifying quality-enhancing commits, which helped us to identify previously reported changes, the relatively low detection rates of existing tools highlight our contribution in shedding light on overlooked quality-enhancing code change practices. 

\vspace{-1em}
\section{Implications}
\label{sec:Implications}
\subsection{For Practitioners \& Researchers:}
Our systematic evaluation of Python ML systems provides practitioners and researchers with a robust, automated tool designed for assessing the code quality of ML-driven software development. This ultimately contributes to the production of high-quality software. Moreover, we believe that the introduction of PyQu, along with the comprehensive dataset of 1,728 quality-enhancing commits we collected and the taxonomy of 61 change types, lays the groundwork and is an essential resource for practitioners, enabling them to effectively implement targeted code changes to enhance specific quality attributes. Additionally, it provides a robust baseline to help guide future research endeavors aimed at advancing the understanding and management of ML systems evolution.
\vspace{-1em}
\subsection{For Tool Builders \& IDEs Designers:}
PyQu enables IDE designers to seamlessly integrate automatic code quality evaluation into development environments, making it easier to maintain high-quality MLS. Tool builders can further benefit from our dataset of code changes and their associated quality improvements, using it as a practical resource to automate and simplify code enhancement processes, thereby boosting the overall development efficiency of MLS.


\vspace{-0.7em}
\section{Threats To Validity}
\label{sec:ThreatsToValidity}
\textbf{Internal Validity:} How reliable are our tool’s results? The success of our study—and the accuracy of our quality-enhancement mapping—hinges on our tool’s ability to precisely mine commits from the subject corpus. We rely on low-level metrics which, as discussed in \autoref{subsec:formulas}, have proven effective for measuring the QAs we examine. To avoid ambiguity in how these metrics are weighted, we employ ML classifiers—a well-established approach in software engineering research \cite{rivera2015incorporating}. For each quality, we select the top-performing classifier to ensure high accuracy and maintain stability on unseen data with minimal training-testing gap. Finally, we manually verified 1,728 commits labeled as enhanced and achieved a precision of 0.84, further validating the tool’s reliability.\\
\textbf{External Validity:} Do our results generalize? We analyzed 3,340 projects across diverse domains and popularity levels, identifying both \rev{Python-specific and language-agnostic quality changes (e.g., removing unused imports, replacing deprecated APIs), supporting the generalizability of our approach beyond Python-based MLS.} While further exploration may uncover additional improvements, our publicly available tool enables others to apply it in new contexts and discover further patterns. Given the large number of commits identified, we used simple random sampling (95\% confidence and a 5\% margin of error) to ensure representativeness. 

\vspace{-1em}
\section{RELATED WORK}
\label{sec:RelatedWork}
\subsection{Refactoring and Code Change Studies in ML Systems}
Prior studies have investigated code changes and refactoring practices specifically within MLS. Dilhara et al. \cite{Dilhara_Ketkar_Sannidhi_Dig_2022} took a data-driven approach by mining Python ML projects to uncover fine-grained change patterns. Their taxonomy of repetitive changes provides a rich foundation for understanding how ML code evolves. Tang et al. \cite{tang2021empirical} further examined refactorings in ML systems and pointed out that many changes—such as duplicated configuration code or minor changes in algorithm implementations—are unique to ML and can easily slip past general-purpose tools.

Other studies have contributed valuable insights into ML code evolution. 
Zhou et al. \cite{zhou2020harp} introduced HARP, a holistic tool for analyzing refactorings in Python analytics programs.
Nguyen et al. \cite{nguyen2019graph} created CPatMiner to mine fine-grained code change patterns at scale, while 
Silva et al. \cite{silva2021refdiff} and Atwi et al. \cite{pyref} looked at the challenges of refactoring across multi-language repositories—underscoring the difficulties of adapting Java-based tools to 
languages like Python. Kim \cite{kim2020software} discussed how ML code is particularly dynamic and complex.
Finally, Dilhara et al. \cite{dilhara2021understanding} provided a longitudinal perspective on ML system evolution.

While existing studies effectively detect and categorize recurring code changes in ML projects, their focus is largely limited to identifying “what” changes occur. In contrast, our work with PyQu advances this area by quantitatively associating specific code changes with measurable improvements in ML software quality. By directly linking code evolution to quality outcomes, our study addresses a critical gap and offers actionable insights for researchers and practitioners alike.
\vspace{-1em}
\subsection{Code Quality Assessment in ML Systems}
Quality assessment in traditional software engineering relies on established metrics like 
maintainability indexes, and coupling/cohesion measures. Chaparro et al. \cite{Chaparro2014impact} demonstrated that systematic refactoring positively impacts software quality attributes. However, ML systems—especially those written in dynamic languages like Python—bring additional challenges. They rely on numerous external libraries, and often incorporate specialized logic for model training and data preprocessing \cite{sculley2015hidden}. 
Sculley et al. \cite{sculley2015hidden} famously argued that ML systems carry “hidden technical debt” which can undermine both maintainability and performance. Kim \cite{kim2020software} emphasized that even minor code issues in ML can have an impact on model outcomes. More recently, studies by Zhang et al. \cite{zhang2018empirical} and Islam et al. \cite{islam2019deep} have identified common bug patterns—like tensor shape mismatches and precision errors—in deep learning applications, highlighting the risks of suboptimal code.


Most existing quality assessment studies primarily focus on identifying potential issues but rarely measure how specific code modifications enhance software quality. Our approach, implemented in PyQu, addresses this limitation by leveraging low-level software metrics to assess the impact of code changes in Python-based ML systems. By explicitly correlating fine-grained code changes with measurable improvements in different quality attributes such as maintainability, reliability, and modularity, we extend traditional quality assessment techniques into the ML domain and offer actionable guidance for optimizing these complex systems.
\vspace{-1em}
\section{CONCLUSION}
\label{sec:Conclusion}

In this study, we addressed a critical gap in understanding how code changes affect Python-based MLS quality. We conducted a large-scale, empirical investigation on code changes in 3,340 open-source Python MLS. We introduced \toolName\, a novel tool that leverages low-level software metrics to identify quality-enhancing commits with an average accuracy, precision, and recall of 0.84 and 0.85 of average F1 score, demonstrating its effectiveness in identifying quality-enhancing modifications in Python-based MLS.

Furthermore, we identified \changes\ distinct types of quality-enhancing code modifications grouped into 13 categories and mapped to the QAs they enhance, namely understandability, maintainability, reliability, modularity, and usability. Notably, our findings reveal \rev{25} previously undocumented changes, offering significant insights to advance state-of-the-art Python changes detection tools. Our work lays a foundation for future research aiming to expand automated quality assessment tools in the rapidly evolving landscape of MLS.


\bibliographystyle{ACM-Reference-Format}
\bibliography{main}


\begin{thebibliography}{99}


\ifx \showCODEN    \undefined \def \showCODEN     #1{\unskip}     \fi
\ifx \showISBNx    \undefined \def \showISBNx     #1{\unskip}     \fi
\ifx \showISBNxiii \undefined \def \showISBNxiii  #1{\unskip}     \fi
\ifx \showISSN     \undefined \def \showISSN      #1{\unskip}     \fi
\ifx \showLCCN     \undefined \def \showLCCN      #1{\unskip}     \fi
\ifx \shownote     \undefined \def \shownote      #1{#1}          \fi
\ifx \showarticletitle \undefined \def \showarticletitle #1{#1}   \fi
\ifx \showURL      \undefined \def \showURL       {\relax}        \fi
\providecommand\bibfield[2]{#2}
\providecommand\bibinfo[2]{#2}
\providecommand\natexlab[1]{#1}
\providecommand\showeprint[2][]{arXiv:#2}

\bibitem[ast(rees)]%
        {ast}
 \bibinfo{year}{Abstract syntax trees}\natexlab{}.
\newblock
\urldef\tempurl%
\url{https://docs.python.org/3/library/ast.html}
\showURL{%
\tempurl}
\newblock
\shownote{Accessed: 2024-12-15}.


\bibitem[c54(1ac3)]%
        {c541ac34}
 \bibinfo{year}{Code refactoring - analysiscenter/batchflow@c541ac3}\natexlab{}.
\newblock
\urldef\tempurl%
\url{https://github.com/analysiscenter/batchflow/commit/c541ac34}
\showURL{%
\tempurl}
\newblock
\shownote{Accessed: 2025-01-18}.


\bibitem[PEP(ions)]%
        {PEP257}
 \bibinfo{year}{Docstring Conventions}\natexlab{}.
\newblock
\urldef\tempurl%
\url{https://peps.python.org/pep-0257/}
\showURL{%
\tempurl}
\newblock
\shownote{Accessed: 2024-12-15}.


\bibitem[Fla(ake8)]%
        {Flake8}
 \bibinfo{year}{Flake8}\natexlab{}.
\newblock \bibinfo{title}{Your Tool For Style Guide Enforcement — flake8 7.0.0 documentation}.
\newblock
\urldef\tempurl%
\url{https://flake8.pycqa.org/en/latest/}
\showURL{%
\tempurl}
\newblock
\shownote{Accessed: 2024-11-01}.


\bibitem[fla(sors)]%
        {flake8-tensors_2024}
 \bibinfo{year}{Flake8-tensors}\natexlab{}.
\newblock
\urldef\tempurl%
\url{https://pypi.org/project/flake8-tensors/}
\showURL{%
\tempurl}
\newblock
\shownote{Accessed: 2024-11-01}.


\bibitem[MLC(PyQu)]%
        {MLCodeQuality}
 \bibinfo{year}{From Code Changes to Quality Gains: An Empirical Study in Python ML Systems with PyQu}\natexlab{}.
\newblock
\urldef\tempurl%
\url{https://sites.google.com/view/mlcodequiality/home}
\showURL{%
\tempurl}
\newblock
\shownote{Accessed: 2025-02-27}.


\bibitem[163(lity)]%
        {1634994}
 \bibinfo{year}{IEEE Standard Dictionary of Measures of the Software Aspects of Dependability}\natexlab{}.
\newblock \bibinfo{journal}{\emph{IEEE Std 982.1-2005 (Revision of IEEE Std 982.1-1988)}} (\bibinfo{year}{IEEE Standard Dictionary of Measures of the Software Aspects of Dependability}), \bibinfo{pages}{1--41}.
\newblock
\href{https://doi.org/10.1109/IEEESTD.2006.215280}{doi:\nolinkurl{10.1109/IEEESTD.2006.215280}}


\bibitem[610(logy)]%
        {610.12-1990}
 \bibinfo{year}{IEEE Standard Glossary of Software Engineering Terminology}\natexlab{}.
\newblock \bibinfo{journal}{\emph{IEEE Std 610.12-1990}} (\bibinfo{year}{IEEE Standard Glossary of Software Engineering Terminology}), \bibinfo{pages}{1--84}.
\newblock
\href{https://doi.org/10.1109/IEEESTD.1990.101064}{doi:\nolinkurl{10.1109/IEEESTD.1990.101064}}


\bibitem[PEP(tion)]%
        {PEP498}
 \bibinfo{year}{Literal String Interpolation}\natexlab{}.
\newblock
\urldef\tempurl%
\url{https://peps.python.org/pep-0498/}
\showURL{%
\tempurl}
\newblock
\shownote{Accessed: 2024-11-24}.


\bibitem[myp(Mypy)]%
        {mypy_2024}
 \bibinfo{year}{Mypy}\natexlab{}.
\newblock
\urldef\tempurl%
\url{https://pypi.org/project/mypy/}
\showURL{%
\tempurl}
\newblock
\shownote{Accessed: 2024-12-16}.


\bibitem[PEP(EP 8)]%
        {PEP8}
 \bibinfo{year}{PEP 8}\natexlab{}.
\newblock \bibinfo{title}{Style Guide for Python Code}.
\newblock
\urldef\tempurl%
\url{https://peps.python.org/pep-0008/}
\showURL{%
\tempurl}
\newblock
\shownote{Accessed: 2025-01-27}.


\bibitem[pyl(lint)]%
        {pylint_2024}
 \bibinfo{year}{Pylint}\natexlab{}.
\newblock
\urldef\tempurl%
\url{https://pypi.org/project/pylint/}
\showURL{%
\tempurl}
\newblock
\shownote{Accessed: 2024-12-16}.


\bibitem[PyT(ials)]%
        {PyTorchTutorials}
 \bibinfo{year}{PyTorch Tutorials}\natexlab{}.
\newblock
\urldef\tempurl%
\url{https://pytorch.org/tutorials/}
\showURL{%
\tempurl}
\newblock
\shownote{Accessed: 2024-12-15}.


\bibitem[rad(PyPI)]%
        {radon_2023}
 \bibinfo{year}{Radon - PyPI}\natexlab{}.
\newblock
\urldef\tempurl%
\url{https://pypi.org/project/radon/}
\showURL{%
\tempurl}
\newblock
\shownote{Accessed: 2024-12-15}.


\bibitem[c11(b191)]%
        {c11b191}
 \bibinfo{year}{Refactored RPN model · broadinstitute/keras-rcnn@c11b191}\natexlab{}.
\newblock
\urldef\tempurl%
\url{https://github.com/broadinstitute/keras-rcnn/commit/c11b191c}
\showURL{%
\tempurl}
\newblock
\shownote{Accessed: 2025-01-18}.


\bibitem[7f6(808f)]%
        {7f6808f}
 \bibinfo{year}{Refactoring lesson \#7. Will not provide notes for this lesson · hunkim/PyTorchZeroToAll@7f6808f}\natexlab{}.
\newblock
\urldef\tempurl%
\url{https://github.com/hunkim/PyTorchZeroToAll/commit/7f6808fbb}
\showURL{%
\tempurl}
\newblock
\shownote{Accessed: 2025-01-18}.


\bibitem[e01(8c81)]%
        {e018c81}
 \bibinfo{year}{Some refactoring for sequence merge · scikit-bio/scikit-bio@e018c81}\natexlab{}.
\newblock
\urldef\tempurl%
\url{https://github.com/scikit-bio/scikit-bio/commit/e018c816}
\showURL{%
\tempurl}
\newblock
\shownote{Accessed: 2025-01-18}.


\bibitem[Eff(tive)]%
        {EffectiveTensorflow2}
 \bibinfo{year}{TensorFlow Effective}\natexlab{}.
\newblock
\urldef\tempurl%
\url{https://www.tensorflow.org/guide/effective_tf2}
\showURL{%
\tempurl}
\newblock
\shownote{Accessed: 2024-12-22}.


\bibitem[tex(stat)]%
        {textstat_2022}
 \bibinfo{year}{Textstat}\natexlab{}.
\newblock
\urldef\tempurl%
\url{https://pypi.org/project/textstat/}
\showURL{%
\tempurl}
\newblock
\shownote{Accessed: 2024-12-10}.


\bibitem[PEP(intsb)]%
        {PEP483}
 \bibinfo{year}{The Theory of Type Hints}\natexlab{b}.
\newblock
\urldef\tempurl%
\url{https://peps.python.org/pep-0483/}
\showURL{%
\tempurl}
\newblock
\shownote{Accessed: 2024-12-15}.


\bibitem[PEP(intsa)]%
        {PEP484}
 \bibinfo{year}{Type Hints}\natexlab{a}.
\newblock
\urldef\tempurl%
\url{https://peps.python.org/pep-0484/}
\showURL{%
\tempurl}
\newblock
\shownote{Accessed: 2025-06-28}.


\bibitem[Aljedaani et~al\mbox{.}(2024)]%
        {aljedaani2024boring}
\bibfield{author}{\bibinfo{person}{Wajdi Aljedaani}, \bibinfo{person}{Anwar Ghammam}, \bibinfo{person}{Mohamed~Wiem Mkaouer}, {and} \bibinfo{person}{Marouane Kessentini}.} \bibinfo{year}{2024}\natexlab{}.
\newblock \showarticletitle{From boring to boarding: Transforming refactoring education with game-based learning}. In \bibinfo{booktitle}{\emph{Proceedings of the ACM/IEEE 8th International Workshop on Games and Software Engineering}}. \bibinfo{pages}{20--27}.
\newblock


\bibitem[Allamanis et~al\mbox{.}(2018)]%
        {allamanis2018survey}
\bibfield{author}{\bibinfo{person}{Miltiadis Allamanis}, \bibinfo{person}{Earl~T Barr}, \bibinfo{person}{Premkumar Devanbu}, {and} \bibinfo{person}{Charles Sutton}.} \bibinfo{year}{2018}\natexlab{}.
\newblock \showarticletitle{A survey of machine learning for big code and naturalness}.
\newblock \bibinfo{journal}{\emph{ACM Computing Surveys (CSUR)}} \bibinfo{volume}{51}, \bibinfo{number}{4} (\bibinfo{year}{2018}), \bibinfo{pages}{1--37}.
\newblock


\bibitem[Athanasopoulos et~al\mbox{.}(2014)]%
        {athanasopoulos2015coupling}
\bibfield{author}{\bibinfo{person}{Dionysis Athanasopoulos}, \bibinfo{person}{Apostolos~V Zarras}, \bibinfo{person}{George Miskos}, \bibinfo{person}{Valerie Issarny}, {and} \bibinfo{person}{Panos Vassiliadis}.} \bibinfo{year}{2014}\natexlab{}.
\newblock \showarticletitle{Cohesion-driven decomposition of service interfaces without access to source code}.
\newblock \bibinfo{journal}{\emph{IEEE Transactions on Services Computing}} \bibinfo{volume}{8}, \bibinfo{number}{4} (\bibinfo{year}{2014}), \bibinfo{pages}{550--562}.
\newblock


\bibitem[Atwi et~al\mbox{.}(2021)]%
        {pyref}
\bibfield{author}{\bibinfo{person}{Hassan Atwi}, \bibinfo{person}{Bin Lin}, \bibinfo{person}{Nikolaos Tsantalis}, \bibinfo{person}{Yutaro Kashiwa}, \bibinfo{person}{Yasutaka Kamei}, \bibinfo{person}{Naoyasu Ubayashi}, \bibinfo{person}{Gabriele Bavota}, {and} \bibinfo{person}{Michele Lanza}.} \bibinfo{year}{2021}\natexlab{}.
\newblock \showarticletitle{PYREF: Refactoring Detection in Python Projects}. In \bibinfo{booktitle}{\emph{2021 IEEE 21st International Working Conference on Source Code Analysis and Manipulation (SCAM)}}. \bibinfo{pages}{136--141}.
\newblock
\href{https://doi.org/10.1109/SCAM52516.2021.00025}{doi:\nolinkurl{10.1109/SCAM52516.2021.00025}}


\bibitem[Baltes and Ralph(2022)]%
        {baltes2022sampling}
\bibfield{author}{\bibinfo{person}{Sebastian Baltes} {and} \bibinfo{person}{Paul Ralph}.} \bibinfo{year}{2022}\natexlab{}.
\newblock \showarticletitle{Sampling in software engineering research: A critical review and guidelines}.
\newblock \bibinfo{journal}{\emph{Empirical Software Engineering}} \bibinfo{volume}{27}, \bibinfo{number}{4} (\bibinfo{year}{2022}), \bibinfo{pages}{94}.
\newblock


\bibitem[Behnamghader et~al\mbox{.}(2017)]%
        {behnamghader2017better}
\bibfield{author}{\bibinfo{person}{Pooyan Behnamghader}, \bibinfo{person}{Reem Alfayez}, \bibinfo{person}{Kamonphop Srisopha}, {and} \bibinfo{person}{Barry Boehm}.} \bibinfo{year}{2017}\natexlab{}.
\newblock \showarticletitle{Towards Better Understanding of Software Quality Evolution through Commit-Impact Analysis}. In \bibinfo{booktitle}{\emph{2017 IEEE International Conference on Software Quality, Reliability and Security (QRS)}}. \bibinfo{pages}{251--262}.
\newblock
\href{https://doi.org/10.1109/QRS.2017.36}{doi:\nolinkurl{10.1109/QRS.2017.36}}


\bibitem[Bhatia and Malhotra(2014)]%
        {bhatia2014loc}
\bibfield{author}{\bibinfo{person}{Sonam Bhatia} {and} \bibinfo{person}{Jyoteesh Malhotra}.} \bibinfo{year}{2014}\natexlab{}.
\newblock \showarticletitle{A survey on impact of lines of code on software complexity}. In \bibinfo{booktitle}{\emph{2014 International Conference on Advances in Engineering \& Technology Research (ICAETR-2014)}}. IEEE, \bibinfo{pages}{1--4}.
\newblock


\bibitem[Blagus and Lusa(2013)]%
        {blagus2013smote}
\bibfield{author}{\bibinfo{person}{Rok Blagus} {and} \bibinfo{person}{Lara Lusa}.} \bibinfo{year}{2013}\natexlab{}.
\newblock \showarticletitle{SMOTE for high-dimensional class-imbalanced data}.
\newblock \bibinfo{journal}{\emph{BMC bioinformatics}}  \bibinfo{volume}{14} (\bibinfo{year}{2013}), \bibinfo{pages}{1--16}.
\newblock


\bibitem[Braiek et~al\mbox{.}(2018)]%
        {braiek2018open}
\bibfield{author}{\bibinfo{person}{Houssem~Ben Braiek}, \bibinfo{person}{Foutse Khomh}, {and} \bibinfo{person}{Bram Adams}.} \bibinfo{year}{2018}\natexlab{}.
\newblock \showarticletitle{The open-closed principle of modern machine learning frameworks}. In \bibinfo{booktitle}{\emph{Proceedings of the 15th international conference on mining software repositories}}. \bibinfo{pages}{353--363}.
\newblock


\bibitem[Campbell et~al\mbox{.}(2013)]%
        {Campbell_Quincy_Osserman_Pedersen_2013}
\bibfield{author}{\bibinfo{person}{John~L. Campbell}, \bibinfo{person}{Charles Quincy}, \bibinfo{person}{Jordan Osserman}, {and} \bibinfo{person}{Ove~K. Pedersen}.} \bibinfo{year}{2013}\natexlab{}.
\newblock \showarticletitle{Coding In-depth Semistructured Interviews}.
\newblock \bibinfo{journal}{\emph{Sociological Methods \& Research}} \bibinfo{volume}{42}, \bibinfo{number}{3} (\bibinfo{date}{Aug.} \bibinfo{year}{2013}), \bibinfo{pages}{294–320}.
\newblock
\href{https://doi.org/10.1177/0049124113500475}{doi:\nolinkurl{10.1177/0049124113500475}}


\bibitem[Chaparro et~al\mbox{.}(2014)]%
        {Chaparro2014impact}
\bibfield{author}{\bibinfo{person}{Oscar Chaparro}, \bibinfo{person}{Gabriele Bavota}, \bibinfo{person}{Andrian Marcus}, {and} \bibinfo{person}{Massimiliano~Di Penta}.} \bibinfo{year}{2014}\natexlab{}.
\newblock \showarticletitle{On the Impact of Refactoring Operations on Code Quality Metrics}. In \bibinfo{booktitle}{\emph{2014 IEEE International Conference on Software Maintenance and Evolution}}. \bibinfo{pages}{456--460}.
\newblock
\href{https://doi.org/10.1109/ICSME.2014.73}{doi:\nolinkurl{10.1109/ICSME.2014.73}}


\bibitem[Chen et~al\mbox{.}(2019)]%
        {chen2019assessing}
\bibfield{author}{\bibinfo{person}{Celia Chen}, \bibinfo{person}{Michael Shoga}, \bibinfo{person}{Brian Li}, {and} \bibinfo{person}{Barry Boehm}.} \bibinfo{year}{2019}\natexlab{}.
\newblock \showarticletitle{Assessing software understandability in systems by leveraging fuzzy method and linguistic analysis}.
\newblock \bibinfo{journal}{\emph{Procedia Computer Science}}  \bibinfo{volume}{153} (\bibinfo{year}{2019}), \bibinfo{pages}{17--26}.
\newblock


\bibitem[Chidamber and Kemerer(1994)]%
        {chidamber1994metrics}
\bibfield{author}{\bibinfo{person}{Shyam~R. Chidamber} {and} \bibinfo{person}{Chris~F. Kemerer}.} \bibinfo{year}{1994}\natexlab{}.
\newblock \showarticletitle{A Metrics Suite for Object Oriented Design}.
\newblock \bibinfo{journal}{\emph{IEEE Transactions on Software Engineering}} \bibinfo{volume}{20}, \bibinfo{number}{6} (\bibinfo{year}{1994}), \bibinfo{pages}{476--493}.
\newblock


\bibitem[Cochran(1977)]%
        {Cochran1977}
\bibfield{author}{\bibinfo{person}{William~Gemmell Cochran}.} \bibinfo{year}{1977}\natexlab{}.
\newblock \bibinfo{booktitle}{\emph{Sampling techniques}}.
\newblock \bibinfo{publisher}{John Wiley \& Sons}.
\newblock


\bibitem[C{\^o}t{\'e} et~al\mbox{.}(2024)]%
        {cote2024quality}
\bibfield{author}{\bibinfo{person}{Pierre-Olivier C{\^o}t{\'e}}, \bibinfo{person}{Amin Nikanjam}, \bibinfo{person}{Rached Bouchoucha}, \bibinfo{person}{Ilan Basta}, \bibinfo{person}{Mouna Abidi}, {and} \bibinfo{person}{Foutse Khomh}.} \bibinfo{year}{2024}\natexlab{}.
\newblock \showarticletitle{Quality issues in machine learning software systems}.
\newblock \bibinfo{journal}{\emph{Empirical Software Engineering}} \bibinfo{volume}{29}, \bibinfo{number}{6} (\bibinfo{year}{2024}), \bibinfo{pages}{149}.
\newblock


\bibitem[Deepset-Ai(2025)]%
        {a5bda34}
\bibfield{author}{\bibinfo{person}{Deepset-Ai}.} \bibinfo{year}{Fix loading of saved models with class weights (\#431) · deepset-ai/FARM@a5bda34 2025}\natexlab{}.
\newblock
\urldef\tempurl%
\url{https://github.com/deepset-ai/FARM/commit/a5bda34f}
\showURL{%
\tempurl}
\newblock
\shownote{Accessed: 2025-01-18}.


\bibitem[DeYoung et~al\mbox{.}(1982)]%
        {DeYoung1982Analyzer-generated}
\bibfield{author}{\bibinfo{person}{Gerrit~E. DeYoung}, \bibinfo{person}{G. Kampen}, {and} \bibinfo{person}{J. Topolski}.} \bibinfo{year}{1982}\natexlab{}.
\newblock \showarticletitle{Analyzer-generated and human-judged predictors of computer program readability}.
\newblock  (\bibinfo{year}{1982}), \bibinfo{pages}{223--228}.
\newblock
\href{https://doi.org/10.1145/800049.801784}{doi:\nolinkurl{10.1145/800049.801784}}


\bibitem[Dheepak and Vaishali(2021)]%
        {dheepak2021comprehensive}
\bibfield{author}{\bibinfo{person}{G Dheepak} {and} \bibinfo{person}{D Vaishali}.} \bibinfo{year}{2021}\natexlab{}.
\newblock \showarticletitle{A comprehensive overview of machine learning algorithms and their applications}.
\newblock \bibinfo{journal}{\emph{International Journal of Advanced Research in Science, Communication and Technology}} (\bibinfo{year}{2021}), \bibinfo{pages}{12--23}.
\newblock


\bibitem[Dilhara et~al\mbox{.}(2021)]%
        {dilhara2021understanding}
\bibfield{author}{\bibinfo{person}{Malinda Dilhara}, \bibinfo{person}{Ameya Ketkar}, {and} \bibinfo{person}{Danny Dig}.} \bibinfo{year}{2021}\natexlab{}.
\newblock \showarticletitle{Understanding software-2.0: A study of machine learning library usage and evolution}.
\newblock \bibinfo{journal}{\emph{ACM Transactions on Software Engineering and Methodology (TOSEM)}} \bibinfo{volume}{30}, \bibinfo{number}{4} (\bibinfo{year}{2021}), \bibinfo{pages}{1--42}.
\newblock


\bibitem[Dilhara et~al\mbox{.}(2022)]%
        {Dilhara_Ketkar_Sannidhi_Dig_2022}
\bibfield{author}{\bibinfo{person}{Malinda Dilhara}, \bibinfo{person}{Ameya Ketkar}, \bibinfo{person}{Nikhith Sannidhi}, {and} \bibinfo{person}{Danny Dig}.} \bibinfo{year}{2022}\natexlab{}.
\newblock \showarticletitle{Discovering repetitive code changes in python ML systems}.
\newblock \bibinfo{journal}{\emph{Proceedings of the 44th International Conference on Software Engineering}} (\bibinfo{date}{May} \bibinfo{year}{2022}).
\newblock
\href{https://doi.org/10.1145/3510003.3510225}{doi:\nolinkurl{10.1145/3510003.3510225}}


\bibitem[Dixon et~al\mbox{.}(2020)]%
        {dixon2020machine}
\bibfield{author}{\bibinfo{person}{Matthew~F Dixon}, \bibinfo{person}{Igor Halperin}, {and} \bibinfo{person}{Paul Bilokon}.} \bibinfo{year}{2020}\natexlab{}.
\newblock \bibinfo{booktitle}{\emph{Machine learning in finance}}. Vol.~\bibinfo{volume}{1170}.
\newblock \bibinfo{publisher}{Springer}.
\newblock


\bibitem[Ebert et~al\mbox{.}(2016)]%
        {ebert2016cyclomatic}
\bibfield{author}{\bibinfo{person}{Christof Ebert}, \bibinfo{person}{James Cain}, \bibinfo{person}{Giuliano Antoniol}, \bibinfo{person}{Steve Counsell}, {and} \bibinfo{person}{Phillip Laplante}.} \bibinfo{year}{2016}\natexlab{}.
\newblock \showarticletitle{Cyclomatic complexity}.
\newblock \bibinfo{journal}{\emph{IEEE software}} \bibinfo{volume}{33}, \bibinfo{number}{6} (\bibinfo{year}{2016}), \bibinfo{pages}{27--29}.
\newblock


\bibitem[Fawcett(2006)]%
        {fawcett2006introduction}
\bibfield{author}{\bibinfo{person}{Tom Fawcett}.} \bibinfo{year}{2006}\natexlab{}.
\newblock \showarticletitle{An Introduction to ROC Analysis}.
\newblock \bibinfo{journal}{\emph{Pattern Recognition Letters}} \bibinfo{volume}{27}, \bibinfo{number}{8} (\bibinfo{year}{2006}), \bibinfo{pages}{861--874}.
\newblock


\bibitem[Fenton and Bieman(2014)]%
        {fenton2014software}
\bibfield{author}{\bibinfo{person}{Norman~E Fenton} {and} \bibinfo{person}{James Bieman}.} \bibinfo{year}{2014}\natexlab{}.
\newblock \bibinfo{booktitle}{\emph{Software metrics: A rigorous and practical approach}}.
\newblock \bibinfo{publisher}{CRC Press}.
\newblock


\bibitem[Fern{\'a}ndez et~al\mbox{.}(2018)]%
        {fernandez2018smote}
\bibfield{author}{\bibinfo{person}{Alberto Fern{\'a}ndez}, \bibinfo{person}{Salvador Garcia}, \bibinfo{person}{Francisco Herrera}, {and} \bibinfo{person}{Nitesh~V Chawla}.} \bibinfo{year}{2018}\natexlab{}.
\newblock \showarticletitle{SMOTE for learning from imbalanced data: progress and challenges, marking the 15-year anniversary}.
\newblock \bibinfo{journal}{\emph{Journal of artificial intelligence research}}  \bibinfo{volume}{61} (\bibinfo{year}{2018}), \bibinfo{pages}{863--905}.
\newblock


\bibitem[Fincher and Tenenberg(2005)]%
        {fincher2005making}
\bibfield{author}{\bibinfo{person}{Sally Fincher} {and} \bibinfo{person}{Josh Tenenberg}.} \bibinfo{year}{2005}\natexlab{}.
\newblock \showarticletitle{Making sense of card sorting data}.
\newblock \bibinfo{journal}{\emph{Expert Systems}} \bibinfo{volume}{22}, \bibinfo{number}{3} (\bibinfo{year}{2005}), \bibinfo{pages}{89--93}.
\newblock


\bibitem[Garnaik et~al\mbox{.}(2014)]%
        {garnaik2014reliability}
\bibfield{author}{\bibinfo{person}{Shrija Garnaik}, \bibinfo{person}{Tirthankar Gayen}, {and} \bibinfo{person}{Samaresh Mishra}.} \bibinfo{year}{2014}\natexlab{}.
\newblock \showarticletitle{Reliability enhancement of software programs by minimizing the overflow errors}.
\newblock \bibinfo{journal}{\emph{International Journal of System Assurance Engineering and Management}}  \bibinfo{volume}{5} (\bibinfo{year}{2014}), \bibinfo{pages}{724--730}.
\newblock


\bibitem[Ghammam et~al\mbox{.}(2025)]%
        {ghammam2025build}
\bibfield{author}{\bibinfo{person}{Anwar Ghammam}, \bibinfo{person}{Dhia~Elhaq Rzig}, \bibinfo{person}{Mohamed Almukhtar}, \bibinfo{person}{Rania Khalsi}, \bibinfo{person}{Foyzul Hassan}, {and} \bibinfo{person}{Marouane Kessentini}.} \bibinfo{year}{2025}\natexlab{}.
\newblock \showarticletitle{Build Code Needs Maintenance Too: A Study on Refactoring and Technical Debt in Build Systems}. In \bibinfo{booktitle}{\emph{2025 IEEE/ACM 22nd International Conference on Mining Software Repositories (MSR)}}. IEEE, \bibinfo{pages}{616--628}.
\newblock


\bibitem[Habehh and Gohel(2021)]%
        {habehh2021machine}
\bibfield{author}{\bibinfo{person}{Hafsa Habehh} {and} \bibinfo{person}{Suril Gohel}.} \bibinfo{year}{2021}\natexlab{}.
\newblock \showarticletitle{Machine learning in healthcare}.
\newblock \bibinfo{journal}{\emph{Current genomics}} \bibinfo{volume}{22}, \bibinfo{number}{4} (\bibinfo{year}{2021}), \bibinfo{pages}{291--300}.
\newblock


\bibitem[Hariprasad et~al\mbox{.}(2017)]%
        {hariprasad2017style}
\bibfield{author}{\bibinfo{person}{T Hariprasad}, \bibinfo{person}{G Vidhyagaran}, \bibinfo{person}{K Seenu}, {and} \bibinfo{person}{Chandrasegar Thirumalai}.} \bibinfo{year}{2017}\natexlab{}.
\newblock \showarticletitle{Software complexity analysis using halstead metrics}. In \bibinfo{booktitle}{\emph{2017 international conference on trends in electronics and informatics (ICEI)}}. IEEE, \bibinfo{pages}{1109--1113}.
\newblock


\bibitem[Harzevili et~al\mbox{.}(2023)]%
        {ShiriHarzevili2022}
\bibfield{author}{\bibinfo{person}{Nima~Shiri Harzevili}, \bibinfo{person}{Jiho Shin}, \bibinfo{person}{Junjie Wang}, \bibinfo{person}{Song Wang}, {and} \bibinfo{person}{Nachiappan Nagappan}.} \bibinfo{year}{2023}\natexlab{}.
\newblock \showarticletitle{Characterizing and understanding software security vulnerabilities in machine learning libraries}. In \bibinfo{booktitle}{\emph{2023 IEEE/ACM 20th International Conference on Mining Software Repositories (MSR)}}. IEEE, \bibinfo{pages}{27--38}.
\newblock


\bibitem[Hellendoorn and Devanbu(2017)]%
        {hellendoorn2017deep}
\bibfield{author}{\bibinfo{person}{Vincent~J Hellendoorn} {and} \bibinfo{person}{Premkumar Devanbu}.} \bibinfo{year}{2017}\natexlab{}.
\newblock \showarticletitle{Are deep neural networks the best choice for modeling source code?}. In \bibinfo{booktitle}{\emph{Proceedings of the 2017 11th Joint meeting on foundations of software engineering}}. \bibinfo{pages}{763--773}.
\newblock


\bibitem[Hoffer et~al\mbox{.}(2017)]%
        {Hoffer2017Train}
\bibfield{author}{\bibinfo{person}{Elad Hoffer}, \bibinfo{person}{Itay Hubara}, {and} \bibinfo{person}{Daniel Soudry}.} \bibinfo{year}{2017}\natexlab{}.
\newblock \showarticletitle{Train longer, generalize better: closing the generalization gap in large batch training of neural networks}.
\newblock \bibinfo{journal}{\emph{ArXiv}}  \bibinfo{volume}{abs/1705.08741} (\bibinfo{year}{2017}).
\newblock


\bibitem[Islam et~al\mbox{.}(2019)]%
        {islam2019deep}
\bibfield{author}{\bibinfo{person}{M.~J. Islam}, \bibinfo{person}{G. Nguyen}, \bibinfo{person}{R. Pan}, {and} \bibinfo{person}{H. Rajan}.} \bibinfo{year}{2019}\natexlab{}.
\newblock \showarticletitle{A Comprehensive Study on Deep Learning Bug Characteristics}. In \bibinfo{booktitle}{\emph{Proceedings of the 41st International Conference on Software Engineering (ICSE '19)}}. \bibinfo{pages}{510--520}.
\newblock
\href{https://doi.org/10.1145/3338906.3338955}{doi:\nolinkurl{10.1145/3338906.3338955}}


\bibitem[Ivers et~al\mbox{.}(2024)]%
        {ivers2024mind}
\bibfield{author}{\bibinfo{person}{James Ivers}, \bibinfo{person}{Anwar Ghammam}, \bibinfo{person}{Khouloud Gaaloul}, \bibinfo{person}{Ipek Ozkaya}, \bibinfo{person}{Marouane Kessentini}, {and} \bibinfo{person}{Wajdi Aljedaani}.} \bibinfo{year}{2024}\natexlab{}.
\newblock \showarticletitle{Mind the gap: The disconnect between refactoring criteria used in industry and refactoring recommendation tools}. In \bibinfo{booktitle}{\emph{2024 IEEE International Conference on Software Maintenance and Evolution (ICSME)}}. IEEE, \bibinfo{pages}{138--150}.
\newblock


\bibitem[Jabborov et~al\mbox{.}(2023)]%
        {jabborov2023taxonomy}
\bibfield{author}{\bibinfo{person}{Ahror Jabborov}, \bibinfo{person}{Arina Kharlamova}, \bibinfo{person}{Zamira Kholmatova}, \bibinfo{person}{Artem Kruglov}, \bibinfo{person}{Vasily Kruglov}, {and} \bibinfo{person}{Giancarlo Succi}.} \bibinfo{year}{2023}\natexlab{}.
\newblock \showarticletitle{Taxonomy of Quality Assessment for Intelligent Software Systems: A Systematic Literature Review}.
\newblock \bibinfo{journal}{\emph{IEEE Access}}  \bibinfo{volume}{11} (\bibinfo{year}{2023}), \bibinfo{pages}{130491--130507}.
\newblock


\bibitem[Khan and Nadeem(2023)]%
        {khan2023apifc}
\bibfield{author}{\bibinfo{person}{Bilal Khan} {and} \bibinfo{person}{Aamer Nadeem}.} \bibinfo{year}{2023}\natexlab{}.
\newblock \showarticletitle{Evaluating the effectiveness of decomposed Halstead Metrics in software fault prediction}.
\newblock \bibinfo{journal}{\emph{PeerJ Computer Science}}  \bibinfo{volume}{9} (\bibinfo{year}{2023}), \bibinfo{pages}{e1647}.
\newblock


\bibitem[Kim(2020)]%
        {kim2020software}
\bibfield{author}{\bibinfo{person}{M. Kim}.} \bibinfo{year}{2020}\natexlab{}.
\newblock \showarticletitle{Software Engineering for Data Analytics}.
\newblock \bibinfo{journal}{\emph{IEEE Software}} \bibinfo{volume}{37}, \bibinfo{number}{4} (\bibinfo{year}{2020}), \bibinfo{pages}{36--42}.
\newblock
\href{https://doi.org/10.1109/MS.2020.2985775}{doi:\nolinkurl{10.1109/MS.2020.2985775}}


\bibitem[Maldil(iner)]%
        {PyRefMiner}
\bibfield{author}{\bibinfo{person}{Maldil}.} \bibinfo{year}{GitHub - maldil/RefactoringMiner}\natexlab{}.
\newblock
\urldef\tempurl%
\url{https://github.com/maldil/RefactoringMiner}
\showURL{%
\tempurl}
\newblock
\shownote{Accessed: 2025-01-15}.


\bibitem[McCabe(1976)]%
        {mccabe1976complexity}
\bibfield{author}{\bibinfo{person}{Thomas~J. McCabe}.} \bibinfo{year}{1976}\natexlab{}.
\newblock \showarticletitle{A Complexity Measure}.
\newblock \bibinfo{journal}{\emph{IEEE Transactions on Software Engineering}} \bibinfo{volume}{SE-2}, \bibinfo{number}{4} (\bibinfo{year}{1976}), \bibinfo{pages}{308--320}.
\newblock


\bibitem[Nguyen et~al\mbox{.}(2019)]%
        {nguyen2019graph}
\bibfield{author}{\bibinfo{person}{Hoan~Anh Nguyen}, \bibinfo{person}{Tien~N Nguyen}, \bibinfo{person}{Danny Dig}, \bibinfo{person}{Son Nguyen}, \bibinfo{person}{Hieu Tran}, {and} \bibinfo{person}{Michael Hilton}.} \bibinfo{year}{2019}\natexlab{}.
\newblock \showarticletitle{Graph-based mining of in-the-wild, fine-grained, semantic code change patterns}. In \bibinfo{booktitle}{\emph{2019 IEEE/ACM 41st International Conference on Software Engineering (ICSE)}}. IEEE, \bibinfo{pages}{819--830}.
\newblock


\bibitem[Nguyen(2020)]%
        {Nguyen2020}
\bibfield{author}{\bibinfo{person}{N~Rich Nguyen}.} \bibinfo{year}{2020}\natexlab{}.
\newblock \showarticletitle{Toward an open-source toolkit for machine learning education}. In \bibinfo{booktitle}{\emph{Proceedings of the 51st ACM Technical Symposium on Computer Science Education}}. \bibinfo{pages}{1400--1400}.
\newblock


\bibitem[Noei et~al\mbox{.}(2024)]%
        {noei2024detecting}
\bibfield{author}{\bibinfo{person}{Shayan Noei}, \bibinfo{person}{Heng Li}, {and} \bibinfo{person}{Ying Zou}.} \bibinfo{year}{2024}\natexlab{}.
\newblock \showarticletitle{Detecting Refactoring Commits in Machine Learning Python Projects: A Machine Learning-Based Approach}.
\newblock \bibinfo{journal}{\emph{ACM Transactions on Software Engineering and Methodology}} (\bibinfo{year}{2024}).
\newblock


\bibitem[Pandey et~al\mbox{.}(2024)]%
        {pandey2024transforming}
\bibfield{author}{\bibinfo{person}{Ruchika Pandey}, \bibinfo{person}{Prabhat Singh}, \bibinfo{person}{Raymond Wei}, {and} \bibinfo{person}{Shaila Shankar}.} \bibinfo{year}{2024}\natexlab{}.
\newblock \showarticletitle{Transforming Software Development: Evaluating the Efficiency and Challenges of GitHub Copilot in Real-World Projects}.
\newblock \bibinfo{journal}{\emph{arXiv preprint arXiv:2406.17910}} (\bibinfo{year}{2024}).
\newblock


\bibitem[Parnas(1972)]%
        {parnas1972criteria}
\bibfield{author}{\bibinfo{person}{David~L. Parnas}.} \bibinfo{year}{1972}\natexlab{}.
\newblock \showarticletitle{On the Criteria to Be Used in Decomposing Systems into Modules}.
\newblock \bibinfo{journal}{\emph{Commun. ACM}} \bibinfo{volume}{15}, \bibinfo{number}{12} (\bibinfo{year}{1972}), \bibinfo{pages}{1053--1058}.
\newblock


\bibitem[Pl{\"o}sch et~al\mbox{.}(2014)]%
        {plösch2014documentation}
\bibfield{author}{\bibinfo{person}{Reinhold Pl{\"o}sch}, \bibinfo{person}{Andreas Dautovic}, {and} \bibinfo{person}{Matthias Saft}.} \bibinfo{year}{2014}\natexlab{}.
\newblock \showarticletitle{The value of software documentation quality}. In \bibinfo{booktitle}{\emph{2014 14th International Conference on Quality Software}}. IEEE, \bibinfo{pages}{333--342}.
\newblock


\bibitem[Posnett et~al\mbox{.}(2011)]%
        {Posnett2011A}
\bibfield{author}{\bibinfo{person}{Daryl Posnett}, \bibinfo{person}{Abram Hindle}, {and} \bibinfo{person}{Premkumar~T. Devanbu}.} \bibinfo{year}{2011}\natexlab{}.
\newblock \showarticletitle{A simpler model of software readability}.
\newblock  (\bibinfo{year}{2011}), \bibinfo{pages}{73--82}.
\newblock
\href{https://doi.org/10.1145/1985441.1985454}{doi:\nolinkurl{10.1145/1985441.1985454}}


\bibitem[Pradipta et~al\mbox{.}(2021)]%
        {Pradipta2021SMOTE}
\bibfield{author}{\bibinfo{person}{G.~A. Pradipta}, \bibinfo{person}{Retantyo Wardoyo}, \bibinfo{person}{Aina Musdholifah}, \bibinfo{person}{I. Sanjaya}, {and} \bibinfo{person}{Muhammad Ismail}.} \bibinfo{year}{2021}\natexlab{}.
\newblock \showarticletitle{SMOTE for Handling Imbalanced Data Problem : A Review}.
\newblock \bibinfo{journal}{\emph{2021 Sixth International Conference on Informatics and Computing (ICIC)}} (\bibinfo{year}{2021}), \bibinfo{pages}{1--8}.
\newblock
\href{https://doi.org/10.1109/ICIC54025.2021.9632912}{doi:\nolinkurl{10.1109/ICIC54025.2021.9632912}}


\bibitem[PyTorch(uide)]%
        {PyTorch}
\bibfield{author}{\bibinfo{person}{Team PyTorch}.} \bibinfo{year}{Pytorch 0.4.0 migration guide}\natexlab{}.
\newblock
\urldef\tempurl%
\url{https://pytorch.org/blog/pytorch-0_4_0-migration-guide/}
\showURL{%
\tempurl}


\bibitem[Quyoum et~al\mbox{.}(2010)]%
        {quyoum2010improving}
\bibfield{author}{\bibinfo{person}{Aasia Quyoum}, \bibinfo{person}{Mehraj-Ud-Din Dar}, {and} \bibinfo{person}{SMK Quadri}.} \bibinfo{year}{2010}\natexlab{}.
\newblock \showarticletitle{Improving Software Reliability using Software Engineering Approach- A Review}.
\newblock \bibinfo{journal}{\emph{International Journal of Computer Applications}} \bibinfo{volume}{10}, \bibinfo{number}{5} (\bibinfo{year}{2010}), \bibinfo{pages}{41--47}.
\newblock


\bibitem[Raschka and Mirjalili(2019)]%
        {raschka2019python}
\bibfield{author}{\bibinfo{person}{Sebastian Raschka} {and} \bibinfo{person}{Vahid Mirjalili}.} \bibinfo{year}{2019}\natexlab{}.
\newblock \bibinfo{booktitle}{\emph{Python machine learning: Machine learning and deep learning with Python, scikit-learn, and TensorFlow 2}}.
\newblock \bibinfo{publisher}{Packt publishing ltd}.
\newblock


\bibitem[Rivera et~al\mbox{.}(2015)]%
        {rivera2015incorporating}
\bibfield{author}{\bibinfo{person}{Nicol{\'a}s Rivera}, \bibinfo{person}{Jorge~A Baier}, {and} \bibinfo{person}{Carlos Hern{\'a}ndez}.} \bibinfo{year}{2015}\natexlab{}.
\newblock \showarticletitle{Incorporating weights into real-time heuristic search}.
\newblock \bibinfo{journal}{\emph{Artificial Intelligence}}  \bibinfo{volume}{225} (\bibinfo{year}{2015}), \bibinfo{pages}{1--23}.
\newblock


\bibitem[Sahu and Srivastava(2019)]%
        {sahu2018revisiting}
\bibfield{author}{\bibinfo{person}{Kavita Sahu} {and} \bibinfo{person}{RK Srivastava}.} \bibinfo{year}{2019}\natexlab{}.
\newblock \showarticletitle{Revisiting software reliability}.
\newblock \bibinfo{journal}{\emph{Data Management, Analytics and Innovation: Proceedings of ICDMAI 2018, Volume 1}} (\bibinfo{year}{2019}), \bibinfo{pages}{221--235}.
\newblock


\bibitem[Sarkar et~al\mbox{.}(2018)]%
        {Sarkar2018}
\bibfield{author}{\bibinfo{person}{Dipanjan Sarkar}, \bibinfo{person}{Raghav Bali}, \bibinfo{person}{Tushar Sharma}, \bibinfo{person}{Dipanjan Sarkar}, \bibinfo{person}{Raghav Bali}, {and} \bibinfo{person}{Tushar Sharma}.} \bibinfo{year}{2018}\natexlab{}.
\newblock \showarticletitle{The Python machine learning ecosystem}.
\newblock \bibinfo{journal}{\emph{Practical Machine Learning with Python: A Problem-Solver's Guide to Building Real-World Intelligent Systems}} (\bibinfo{year}{2018}), \bibinfo{pages}{67--118}.
\newblock


\bibitem[Scalabrino et~al\mbox{.}(2017)]%
        {Scalabrino2017}
\bibfield{author}{\bibinfo{person}{Simone Scalabrino}, \bibinfo{person}{Gabriele Bavota}, \bibinfo{person}{Christopher Vendome}, \bibinfo{person}{Mario Linares-V\'{a}squez}, \bibinfo{person}{Denys Poshyvanyk}, {and} \bibinfo{person}{Rocco Oliveto}.} \bibinfo{year}{2017}\natexlab{}.
\newblock \showarticletitle{Automatically assessing code understandability: how far are we?}. In \bibinfo{booktitle}{\emph{Proceedings of the 32nd IEEE/ACM International Conference on Automated Software Engineering}} (Urbana-Champaign, IL, USA) \emph{(\bibinfo{series}{ASE '17})}. \bibinfo{publisher}{IEEE Press}, \bibinfo{pages}{417–427}.
\newblock
\showISBNx{9781538626849}


\bibitem[Scalabrino et~al\mbox{.}(2019)]%
        {Scalabrino2019Automatically}
\bibfield{author}{\bibinfo{person}{Simone Scalabrino}, \bibinfo{person}{G. Bavota}, \bibinfo{person}{Christopher Vendome}, \bibinfo{person}{Mario Linares-Vásquez}, \bibinfo{person}{D. Poshyvanyk}, {and} \bibinfo{person}{Rocco Oliveto}.} \bibinfo{year}{2019}\natexlab{}.
\newblock \showarticletitle{Automatically Assessing Code Understandability}.
\newblock \bibinfo{journal}{\emph{IEEE Transactions on Software Engineering}}  \bibinfo{volume}{47} (\bibinfo{year}{2019}), \bibinfo{pages}{595--613}.
\newblock
\href{https://doi.org/10.1109/TSE.2019.2901468}{doi:\nolinkurl{10.1109/TSE.2019.2901468}}


\bibitem[Scalabrino et~al\mbox{.}(2018)]%
        {Scalabrino_Linares‐Vásquez_Oliveto_Poshyvanyk_2018}
\bibfield{author}{\bibinfo{person}{Simone Scalabrino}, \bibinfo{person}{Mario Linares‐Vásquez}, \bibinfo{person}{Rocco Oliveto}, {and} \bibinfo{person}{Denys Poshyvanyk}.} \bibinfo{year}{2018}\natexlab{}.
\newblock \showarticletitle{A comprehensive model for code readability}.
\newblock \bibinfo{journal}{\emph{Journal of Software}} \bibinfo{volume}{30}, \bibinfo{number}{6} (\bibinfo{date}{June} \bibinfo{year}{2018}).
\newblock
\href{https://doi.org/10.1002/smr.1958}{doi:\nolinkurl{10.1002/smr.1958}}


\bibitem[Scalabrino et~al\mbox{.}(2016)]%
        {scalabrino2016improving}
\bibfield{author}{\bibinfo{person}{Simone Scalabrino}, \bibinfo{person}{M. Vásquez}, \bibinfo{person}{D. Poshyvanyk}, {and} \bibinfo{person}{R. Oliveto}.} \bibinfo{year}{2016}\natexlab{}.
\newblock \showarticletitle{Improving code readability models with textual features}.
\newblock \bibinfo{journal}{\emph{2016 IEEE 24th International Conference on Program Comprehension (ICPC)}} (\bibinfo{year}{2016}), \bibinfo{pages}{1--10}.
\newblock
\href{https://doi.org/10.1109/ICPC.2016.7503707}{doi:\nolinkurl{10.1109/ICPC.2016.7503707}}


\bibitem[Sculley et~al\mbox{.}(2015)]%
        {sculley2015hidden}
\bibfield{author}{\bibinfo{person}{D. Sculley}, \bibinfo{person}{G. Holt}, \bibinfo{person}{D. Golovin}, \bibinfo{person}{E. Davydov}, \bibinfo{person}{T. Phillips}, \bibinfo{person}{D. Ebner}, \bibinfo{person}{V. Chaudhary}, \bibinfo{person}{M. Young}, \bibinfo{person}{J.-F. Crespo}, {and} \bibinfo{person}{D. Dennison}.} \bibinfo{year}{2015}\natexlab{}.
\newblock \showarticletitle{Hidden Technical Debt in Machine Learning Systems}. In \bibinfo{booktitle}{\emph{Advances in Neural Information Processing Systems (NIPS 2015)}}.
\newblock


\bibitem[Shih et~al\mbox{.}(2001)]%
        {shih2001cohesion}
\bibfield{author}{\bibinfo{person}{Timothy Shih}, \bibinfo{person}{Ming-Chi Lee}, \bibinfo{person}{Teh-Sheng Huang}, \bibinfo{person}{Lawrence Deng}, {et~al\mbox{.}}} \bibinfo{year}{2001}\natexlab{}.
\newblock \showarticletitle{Assessing Software Quality Through Visualised Cohesion Metrics}.
\newblock \bibinfo{journal}{\emph{Australasian Journal of Information Systems}} \bibinfo{volume}{8}, \bibinfo{number}{2} (\bibinfo{year}{2001}).
\newblock


\bibitem[Silva et~al\mbox{.}(2021)]%
        {silva2021refdiff}
\bibfield{author}{\bibinfo{person}{Danilo Silva}, \bibinfo{person}{João~Paulo da Silva}, \bibinfo{person}{Gustavo Santos}, \bibinfo{person}{Ricardo Terra}, {and} \bibinfo{person}{Marco~Tulio Valente}.} \bibinfo{year}{2021}\natexlab{}.
\newblock \showarticletitle{RefDiff 2.0: A Multi-Language Refactoring Detection Tool}.
\newblock \bibinfo{journal}{\emph{IEEE Transactions on Software Engineering}} \bibinfo{volume}{47}, \bibinfo{number}{12} (\bibinfo{year}{2021}), \bibinfo{pages}{2786--2802}.
\newblock
\href{https://doi.org/10.1109/TSE.2020.2968072}{doi:\nolinkurl{10.1109/TSE.2020.2968072}}


\bibitem[Sj{\o}berg et~al\mbox{.}(2012)]%
        {sjoberg2012questioning}
\bibfield{author}{\bibinfo{person}{Dag~IK Sj{\o}berg}, \bibinfo{person}{Bente Anda}, {and} \bibinfo{person}{Audris Mockus}.} \bibinfo{year}{2012}\natexlab{}.
\newblock \showarticletitle{Questioning software maintenance metrics: a comparative case study}. In \bibinfo{booktitle}{\emph{Proceedings of the ACM-IEEE international symposium on Empirical software engineering and measurement}}. \bibinfo{pages}{107--110}.
\newblock


\bibitem[Skiada et~al\mbox{.}(2018)]%
        {peggy2018exploring}
\bibfield{author}{\bibinfo{person}{Peggy Skiada}, \bibinfo{person}{Apostolos Ampatzoglou}, \bibinfo{person}{Elvira-Maria Arvanitou}, \bibinfo{person}{Alexander Chatzigeorgiou}, {and} \bibinfo{person}{Ioannis Stamelos}.} \bibinfo{year}{2018}\natexlab{}.
\newblock \showarticletitle{Exploring the Relationship between Software Modularity and Technical Debt}. In \bibinfo{booktitle}{\emph{2018 44th Euromicro Conference on Software Engineering and Advanced Applications (SEAA)}}. \bibinfo{pages}{404--407}.
\newblock
\href{https://doi.org/10.1109/SEAA.2018.00072}{doi:\nolinkurl{10.1109/SEAA.2018.00072}}


\bibitem[{\'S}liwerski et~al\mbox{.}(2005)]%
        {sliwerski2005changes}
\bibfield{author}{\bibinfo{person}{Jacek {\'S}liwerski}, \bibinfo{person}{Thomas Zimmermann}, {and} \bibinfo{person}{Andreas Zeller}.} \bibinfo{year}{2005}\natexlab{}.
\newblock \showarticletitle{When do changes induce fixes?}
\newblock \bibinfo{journal}{\emph{ACM sigsoft software engineering notes}} \bibinfo{volume}{30}, \bibinfo{number}{4} (\bibinfo{year}{2005}), \bibinfo{pages}{1--5}.
\newblock


\bibitem[Sommerville(2011)]%
        {sommerville2011software}
\bibfield{author}{\bibinfo{person}{Ian Sommerville}.} \bibinfo{year}{2011}\natexlab{}.
\newblock \bibinfo{booktitle}{\emph{Software Engineering}}.
\newblock \bibinfo{publisher}{Addison-Wesley}.
\newblock


\bibitem[Spencer and Warfel(2004)]%
        {spencer2004card}
\bibfield{author}{\bibinfo{person}{Donna Spencer} {and} \bibinfo{person}{Todd Warfel}.} \bibinfo{year}{2004}\natexlab{}.
\newblock \showarticletitle{Card sorting: a definitive guide}.
\newblock \bibinfo{journal}{\emph{Boxes and arrows}} \bibinfo{volume}{2}, \bibinfo{number}{2004} (\bibinfo{year}{2004}), \bibinfo{pages}{1--23}.
\newblock


\bibitem[Stevens et~al\mbox{.}(1974)]%
        {stevens1974structured}
\bibfield{author}{\bibinfo{person}{Wayne~P. Stevens}, \bibinfo{person}{Glenford~J. Myers}, {and} \bibinfo{person}{Larry~L. Constantine}.} \bibinfo{year}{1974}\natexlab{}.
\newblock \showarticletitle{Structured Design}.
\newblock \bibinfo{journal}{\emph{IBM Systems Journal}} \bibinfo{volume}{13}, \bibinfo{number}{2} (\bibinfo{year}{1974}), \bibinfo{pages}{115--139}.
\newblock


\bibitem[Tang and Nadi(2023)]%
        {tang2023documentation}
\bibfield{author}{\bibinfo{person}{Henry Tang} {and} \bibinfo{person}{Sarah Nadi}.} \bibinfo{year}{2023}\natexlab{}.
\newblock \showarticletitle{Evaluating Software Documentation Quality}. In \bibinfo{booktitle}{\emph{2023 IEEE/ACM 20th International Conference on Mining Software Repositories (MSR)}}. \bibinfo{pages}{67--78}.
\newblock
\href{https://doi.org/10.1109/MSR59073.2023.00023}{doi:\nolinkurl{10.1109/MSR59073.2023.00023}}


\bibitem[Tang et~al\mbox{.}(2021)]%
        {tang2021empirical}
\bibfield{author}{\bibinfo{person}{Yiming Tang}, \bibinfo{person}{Raffi Khatchadourian}, \bibinfo{person}{Mehdi Bagherzadeh}, \bibinfo{person}{Rhia Singh}, \bibinfo{person}{Ajani Stewart}, {and} \bibinfo{person}{Anita Raja}.} \bibinfo{year}{2021}\natexlab{}.
\newblock \showarticletitle{An empirical study of refactorings and technical debt in machine learning systems}. In \bibinfo{booktitle}{\emph{2021 IEEE/ACM 43rd international conference on software engineering (ICSE)}}. IEEE, \bibinfo{pages}{238--250}.
\newblock


\bibitem[Tensorflow(yone)]%
        {Tensorflow}
\bibfield{author}{\bibinfo{person}{Tensorflow}.} \bibinfo{year}{Tensorflow/tensorflow: An open source machine learning framework for everyone}\natexlab{}.
\newblock
\urldef\tempurl%
\url{https://github.com/tensorflow/tensorflow}
\showURL{%
\tempurl}


\bibitem[Treude et~al\mbox{.}(2020)]%
        {treude2020documentation}
\bibfield{author}{\bibinfo{person}{Christoph Treude}, \bibinfo{person}{Justin Middleton}, {and} \bibinfo{person}{Thushari Atapattu}.} \bibinfo{year}{2020}\natexlab{}.
\newblock \showarticletitle{Beyond accuracy: Assessing software documentation quality}. In \bibinfo{booktitle}{\emph{Proceedings of the 28th ACM Joint Meeting on European Software Engineering Conference and Symposium on the Foundations of Software Engineering}}. \bibinfo{pages}{1509--1512}.
\newblock


\bibitem[Usman et~al\mbox{.}(2017)]%
        {usman2017taxonomies}
\bibfield{author}{\bibinfo{person}{Muhammad Usman}, \bibinfo{person}{Ricardo Britto}, \bibinfo{person}{J{\"u}rgen B{\"o}rstler}, {and} \bibinfo{person}{Emilia Mendes}.} \bibinfo{year}{2017}\natexlab{}.
\newblock \showarticletitle{Taxonomies in software engineering: A systematic mapping study and a revised taxonomy development method}.
\newblock \bibinfo{journal}{\emph{Information and Software Technology}}  \bibinfo{volume}{85} (\bibinfo{year}{2017}), \bibinfo{pages}{43--59}.
\newblock


\bibitem[Venkatkrishna et~al\mbox{.}(2023)]%
        {venkatkrishna2023docgen}
\bibfield{author}{\bibinfo{person}{Vatsal Venkatkrishna}, \bibinfo{person}{Durga~Shree Nagabushanam}, \bibinfo{person}{Emmanuel Iko-Ojo Simon}, {and} \bibinfo{person}{Melina Vidoni}.} \bibinfo{year}{2023}\natexlab{}.
\newblock \showarticletitle{DocGen: Generating Detailed Parameter Docstrings in Python}.
\newblock \bibinfo{journal}{\emph{arXiv preprint arXiv:2311.06453}} (\bibinfo{year}{2023}).
\newblock


\bibitem[Vitale et~al\mbox{.}(2023)]%
        {vitale2023using}
\bibfield{author}{\bibinfo{person}{Antonio Vitale}, \bibinfo{person}{Valentina Piantadosi}, \bibinfo{person}{Simone Scalabrino}, {and} \bibinfo{person}{Rocco Oliveto}.} \bibinfo{year}{2023}\natexlab{}.
\newblock \showarticletitle{Using Deep Learning to Automatically Improve Code Readability}. In \bibinfo{booktitle}{\emph{2023 38th IEEE/ACM International Conference on Automated Software Engineering (ASE)}}. IEEE, \bibinfo{pages}{573--584}.
\newblock


\bibitem[Wen et~al\mbox{.}(2019)]%
        {wen2019}
\bibfield{author}{\bibinfo{person}{Fengcai Wen}, \bibinfo{person}{Csaba Nagy}, \bibinfo{person}{Gabriele Bavota}, {and} \bibinfo{person}{Michele Lanza}.} \bibinfo{year}{2019}\natexlab{}.
\newblock \showarticletitle{A large-scale empirical study on code-comment inconsistencies}. In \bibinfo{booktitle}{\emph{Proceedings of the 27th International Conference on Program Comprehension}} (Montreal, Quebec, Canada) \emph{(\bibinfo{series}{ICPC '19})}. \bibinfo{publisher}{IEEE Press}, \bibinfo{pages}{53–64}.
\newblock
\href{https://doi.org/10.1109/ICPC.2019.00019}{doi:\nolinkurl{10.1109/ICPC.2019.00019}}


\bibitem[Yacouby and Axman(2020)]%
        {Yacouby2020Probabilistic}
\bibfield{author}{\bibinfo{person}{Reda Yacouby} {and} \bibinfo{person}{Dustin Axman}.} \bibinfo{year}{2020}\natexlab{}.
\newblock \showarticletitle{Probabilistic Extension of Precision, Recall, and F1 Score for More Thorough Evaluation of Classification Models}.
\newblock \bibinfo{journal}{\emph{Proceedings of the First Workshop on Evaluation and Comparison of NLP Systems}} (\bibinfo{year}{2020}).
\newblock
\href{https://doi.org/10.18653/v1/2020.eval4nlp-1.9}{doi:\nolinkurl{10.18653/v1/2020.eval4nlp-1.9}}


\bibitem[Zhang et~al\mbox{.}(2018)]%
        {zhang2018empirical}
\bibfield{author}{\bibinfo{person}{Y. Zhang}, \bibinfo{person}{Y. Chen}, \bibinfo{person}{S.-C. Cheung}, \bibinfo{person}{Y. Xiong}, {and} \bibinfo{person}{L. Zhang}.} \bibinfo{year}{2018}\natexlab{}.
\newblock \showarticletitle{An Empirical Study on TensorFlow Program Bugs}. In \bibinfo{booktitle}{\emph{Proceedings of the International Symposium on Software Testing and Analysis}}.
\newblock
\href{https://doi.org/10.1145/3213846.3213866}{doi:\nolinkurl{10.1145/3213846.3213866}}


\bibitem[Zhou et~al\mbox{.}(2020)]%
        {zhou2020harp}
\bibfield{author}{\bibinfo{person}{Weijie Zhou}, \bibinfo{person}{Yue Zhao}, \bibinfo{person}{Guoqiang Zhang}, {and} \bibinfo{person}{Xipeng Shen}.} \bibinfo{year}{2020}\natexlab{}.
\newblock \showarticletitle{HARP: holistic analysis for refactoring Python-based analytics programs}. In \bibinfo{booktitle}{\emph{Proceedings of the ACM/IEEE 42nd International Conference on Software Engineering}}. \bibinfo{pages}{506--517}.
\newblock


\end{thebibliography}

\end{document}